\documentclass[a4paper]{article}
\pdfoutput=1
\usepackage{jheppub}
\usepackage{url}
\usepackage{graphicx}
\usepackage{subcaption}
\usepackage{booktabs}
\usepackage{amsmath}
\usepackage{amssymb}
\usepackage{amstext}
\usepackage{amsfonts}
\usepackage{amsthm}
\usepackage{hyperref}
\usepackage{appendix}
\usepackage{physics}
\usepackage[utf8]{inputenc}
\usepackage{soul}
\usepackage{hhline}
\usepackage{comment}
\usepackage{color}
\usepackage{bm}
\usepackage[normalem]{ulem}
\usepackage{subcaption}
\newcommand{\uvec}[1]{\boldsymbol{\hat{\textbf{#1}}}}

\newcommand\eea{\end{eqnarray}}
\newcommand\bea{\begin{eqnarray}}
\def\vec{\boldsymbol}
\def\l{\left(}
\def\r{\right)}

\title{Searching for Ultralight Scalar Dark Matter with Clocks in Low Earth Orbit}
\date{\today}
 \author[a]{Dawid Brzeminski}
 \author[a]{Aaron Pierce}

\affiliation[a]{
Leinweber Institute for Theoretical Physics, Department of Physics,
University of Michigan, Ann Arbor, MI 48109, USA}

 \emailAdd{dawid@umich.edu}
 \emailAdd{atpierce@umich.edu}
\abstract{The density of ultralight dark matter can be modified in the vicinity of macroscopic bodies when the dark matter possesses quadratic couplings to the Standard Model. If these couplings are sufficiently strong, Earth's atmosphere acts to shield the dark matter, thereby limiting the effectiveness of laboratory-based experiments.
Experiments performed at altitudes exceeding the dark matter de Broglie wavelength experience the same orbit-averaged field amplitude as in the absence of scattering.    
Quantum clocks are capable of detecting variations in fundamental parameters due to the dark matter background.  If based on the International Space Station, they are therefore well-suited to probe dark matter masses  $m_{\rm DM}\gtrsim 10^{-9} \text{\, eV}$.
Moreover, when the dark matter de Broglie wavelength is smaller than Earth's radius ($m_{\rm DM} \gtrsim 10^{-10}$ eV), the dark matter profile around Earth exhibits a dipole feature. In Low Earth Orbits this dipole temporally modulates potential dark matter signals.  This provides a powerful cross-check of the orbit-averaged effect and can enhance the sensitivity of these experiments.  We find optical clocks could give rise to world-leading constraints in some cases. Orbiting nuclear clocks could probe even more of the parameter space inaccessible to ground-based experiments.}

\begin{document}
\preprint{LITP-26-03}
\maketitle
\section{Introduction}

The nature of dark matter (DM) remains elusive. Ultralight dark matter is an interesting possibility; see \cite{Antypas:2022asj} for a review. When it couples linearly to the Standard Model (SM), it induces a Yukawa force and is subject to strong constraints from weak equivalence principle (fifth-force) tests \cite{Wagner:2012ui,Touboul:2017grn,MICROSCOPE:2022doy}. However, if these linear couplings are suppressed, perhaps by a symmetry, new phenomenology from quadratic couplings results \cite{Banerjee:2022sqg}.

This quadratically-coupled dark matter is the focus of this work.   Because macroscopic bodies no longer source the dark matter field, fifth-force bounds do not apply as they do in the linear case. Instead, the couplings to the SM change the effective mass of the DM in the presence of matter. This acts to distort the DM field around macroscopic bodies \cite{Hees:2018fpg}. This distortion gives gradients in a potential that can induce a force which can be probed by fifth-force experiments \cite{Hees:2018fpg, Banerjee:2022sqg, VanTilburg:2024xib, Gan:2025nlu}. 
If couplings to the SM are too strong, the dark matter field reflects from the atmosphere and never reaches the surface. This results in a maximum dark matter-SM coupling that Earth-based experiments can probe. Big Bang Nucleosynthesis (BBN) can, in principle, probe larger couplings \cite{Bouley:2022eer}.  These limits rely on substantial variation of fundamental parameters in the early universe. In models where the scalar is a pseudo-Nambu-Goldstone-boson \cite{Brzeminski:2020uhm, Delaunay:2025pho} and the periodicity of the field limits the effective variation of the coupling, the direct applicability of these bounds is subtle. More generally, the early cosmology of ultralight dark matter is model-dependent. It is of interest to develop bounds that exclusively rely on astrophysics and local experiments. Currently, the best cosmology-independent limits on this otherwise inaccessible parameter space come from the MICROSCOPE experiment \cite{Touboul:2017grn,MICROSCOPE:2022doy, Hees:2018fpg, Banerjee:2022sqg}, a 2.5 year satellite mission that searched for violations of the weak equivalence principle using  differential electrostatic accelerometers. 
However, a large region of parameter space below these MICROSCOPE bounds remains inaccessible to Earth-bound experiments. One way to access this parameter space would be to design a successor to MICROSCOPE, a next-generation space-based fifth-force experiment \cite{Nobili:2012uj}. 

We explore a complementary approach.  We discuss a way to measure the ``potential" instead of measuring the force.  We look for an effect proportional to the density of DM itself, proportional to square of the dark matter field amplitude (the force is the gradient of DM density). As we will review in Sec.~\ref{sec: time-dep fund param}, quadratic couplings to the SM can result in variations in fundamental parameters proportional to the dark matter density \cite{Hees:2018fpg, Stadnik:2020bfk, Banerjee:2022sqg}.  Atomic and nuclear clocks are a particularly promising tool because of their exquisite sensitivity to variations in fundamental parameters \cite{Arvanitaki:2014faa,Safronova:2017xyt,PhysRevA.109.063115, Fuchs:2024xvc,Delaunay:2025lgk}. The parameter space of primary interest to us will be dark matter masses $m_{\rm DM} \gg 3\times 10^{-11} \text{ eV}$ where the de Broglie wavelength is much shorter than Earth's radius, but still light enough that the field description for the dark matter is valid ($m_{\rm DM} \ll$ eV). Previous work demonstrating the sensitivity of space-based clock experiments to quadratic couplings have assumed that the scalar field around Earth is spherically symmetric \cite{Banerjee:2022sqg}.  This assumption applies only when the de Broglie wavelength of the DM is much longer than Earth’s radius $\lambda_{\rm DM}/R_\oplus \gg 1$. 

Quantum clocks typically look for oscillations of  fundamental parameters with a frequency $\omega = 2 m_{\rm DM}$ 
\cite{Hees:2018fpg}, for a review see \cite{Antypas:2022asj}. 
Here, we take advantage of a special feature of quadratic couplings, an effective DC shift of fundamental parameters that depends on the DM density. Even if oscillations in fundamental parameters happen rapidly, it is still possible to place constraints on the dark matter if clocks are placed in locations that have different expectation values for the dark matter density \cite{Stadnik:2020bfk,Masia-Roig:2022net}. Alternately, we can compare clocks with differing architecture as they travel together through a spatially varying background of DM. We provide background on these concepts in Sec.~\ref{sec: time-dep fund param}.  We discuss experimental set-ups for these clock experiments in Sec.~\ref{sec:Space To Ground} and Sec.~\ref{sec:Space to Space}.  We will focus on the case where the sign of the DM-Standard Model couplings are such that the DM mass increases in dense SM media.  For the opposite sign, there is the additional wrinkle of possible resonances -- a decrease in the DM wavelength in the Earth due to the lower DM mass can actually lead to an increase DM density near Earth's surface.  This has been 
discussed in a related context in \cite{Banerjee:2025dlo,delCastillo:2025rbr,Cheng:2025fak,Bartnick:2025lbg}.\footnote{In this case, the detailed density profile of the Earth can be relevant for  reliable calculations.}

Part of the experiment must be performed in space to evade atmospheric shielding. It is natural to ask precisely where the experiment should be deployed to attain optimal sensitivity. Since the DM wind has a preferred direction, the field profile around the Earth can be roughly described as a sum of a monopole and dipole term, which we demonstrate in Sec.~\ref{Sec: matter effect} with a full directional calculation. When the dark matter de Broglie wavelength is much longer than the Earth's radius, the dipole term is suppressed and phenomenologically irrelevant \cite{Hees:2018fpg,Banerjee:2022sqg,Gan:2025nlu}. 
However, in the opposite regime, the dipole feature should no longer be neglected. The presence of the dipole enables modulation of  a clock signal.  This allows for new measurement strategies that can achieve the best overall sensitivity for some DM parameter space. As we show in Sec.~\ref{sec:background}, the dipole is most pronounced at low altitudes, which motivates the use of Low Earth Orbits (LEO). The relevance of LEO is further supported by the ACES mission \cite{Meynadier:2017oha,Savalle:2019isy}, which already operates in the altitude range where dipole effects are enhanced, as well as by future missions involving optical clocks \cite{Shen:2021scn,Schkolnik:2022utn,Xia:2025css}.

The paper is organized as follows. In Sec.~\ref{sec:background}  we discuss pertinent background information, including how the variation of fundamental parameters arises in a DM background and the DM-Earth scattering problem for the quadratically coupled scalar.  A key result of this section is the characterization of the field profile around the Earth. In Sec.~\ref{sec:clocks} we discuss the detection of such a profile using quantum clocks and estimate their sensitivity. We conclude in Sec.~\ref{sec:conclusions}.
\section{Quadratically coupled DM and its interactions with Matter}
\label{sec:background}
In the context of fifth-force searches and the variation of fundamental parameters, it is customary to parametrize couplings to the SM in terms of low-energy dilaton-like couplings \cite{Damour:2010rp,Damour:2010rm, Arvanitaki:2014faa,Banerjee:2022sqg} 
 \bea
 \label{dilaton_couplings}
\mathcal{L} \supset \frac{\kappa^2 \phi^2}{2}\Big[ \frac{d_e^{(2)} }{4 e^2} F_{\mu\nu}F^{\mu\nu}-\frac{d_g^{(2)}\beta_3}{2g_3}G^A_{\mu\nu}G^{A \mu\nu}-d_{m_e}^{(2)}m_e \bar{\psi}_e\psi_e \\
-\sum_{i=u,d}{(d_{m_i}^{(2)}+\gamma_{m_i} d_g^{(2)})m
_i\bar{\psi}_i \psi_i} \Big] \;, \nonumber
\eea
where $\kappa \equiv \sqrt{4\pi G_N}$ with $G_{N}$ Newton's gravitational constant, which acts as a convenient overall normalization; $e$ and $g_3$ are coupling constants of electromagnetism and quantum chromodynamics (QCD), respectively, while $\beta_3 = \partial g_3 / \partial \log \mu$ is the QCD beta function. Here, 
 $F_{\mu \nu}$ and $G_{\mu \nu}$
are the photon and gluon field strengths, respectively. The parameters $m_e$ and $m_{i}$ denote the electron and light quark masses, and $\gamma_{m_i} = - \partial \log m_i/\partial \log \mu$. The couplings $d^{(2)}_x$ with $x\in (g,e,m_e,m_i)$ represent couplings to respective gauge bosons and fermions. Additionally, instead of up- and down-quark masses it is conventional to constrain the couplings to the average quark mass $\hat{m} \equiv (m_u+m_d)/2$ and the quark mass difference $\delta m \equiv m_d-m_u$:
\bea \label{eq: mhat deltam}
d_{\hat{m}} \equiv \frac{m_u d^{(2)}_{m_u}+m_d d^{(2)}_{m_d}}{m_u +m_d}, \qquad d_{\delta m} \equiv \frac{m_d d^{(2)}_{m_d} - m_u d^{(2)}_{m_u}}{m_d - m_u}.
\eea
Some works use an alternate parametrization where couplings are instead expressed as a dimensionful parameter $\Lambda \equiv 1/(\kappa \sqrt{d_{X}^{(2)}})$. We will use the convention from Eqs. \eqref{dilaton_couplings} and \eqref{eq: mhat deltam}.  For examples of how such couplings may be generated, see Refs.~\cite{Banerjee:2022sqg,Gan:2025nlu,Brzeminski:2020uhm, Delaunay:2025pho}.  

\subsection{Dark matter modification of fundamental parameters}  \label{sec: time-dep fund param}
Scalar ultralight dark matter ($m_{\rm DM} \ll$ eV) can be modeled as a classical background field.  A key consequence of this background is a shift in the values of fundamental parameters. For quadratically coupled scalars, fundamental parameters $X$ are sensitive to the square of the field:
\bea
X(\Vec{x},t) = X_0\l 1 + \frac{d_X^{(2)} \phi^2 (\Vec{x},t)}{2} \r,
\eea
where $X$ represents the value of the parameter in the presence of the dark matter field $\phi$ with position- and time-dependent field value $\phi(\Vec{x},t)$.
Since dark matter is non-relativistic, the field oscillates with an angular frequency approximately equal to its mass
\bea
\label{eq: simpleosc}
\phi(\Vec{x},t) \simeq \phi_0(\Vec{x},t) \cos (m_{\rm DM} t + \varphi_0 (t)),
\eea
where $\phi_{0}(\Vec{x},t)$ is the amplitude of DM and $\varphi_0(t)$ is the relative phase. The spatial and temporal dependence of $\phi_{0}(\Vec{x},t)$ is due to a finite coherence length and time. For instance, for time intervals shorter than the DM coherence time $t_{\rm coh} \simeq 1 \, \text{hr}  \l \frac{m_{\rm DM}}{10^{-12}\, \text{eV}} \r^{-1}$ the amplitude and the relative phase at a given location are constant to a good approximation. However, for longer intervals, they undergo $\mathcal{O}(1)$ fluctuations.
In the vacuum the expectation value is $\langle \phi^2(\vec{x},t) \rangle = \frac{\sqrt{\rho}}{m}$, i.e. isotropic.

We are primarily interested in a situation where DM is reflected from the Earth’s surface or the atmosphere, resulting in a location-dependent expectation value $\langle \phi^2(\vec{x},t) \rangle$, which we refer to as the field profile. The $\mathcal{O}(1)$ fluctuations that result for $t > t_{\rm coh}$ act as a source of Gaussian noise.  As we move to long experimental times $T_{exp}$, so that the number of coherence times $N_{\rm coh} = T_{\rm exp}/t_{\rm coh}$ is large, we have $\delta \langle \phi^2 \rangle/ \langle\phi^2 \rangle \sim 1/\sqrt{N_{\rm coh}}$. The typical duration of experiments we consider (weeks to years) is much longer than the DM coherence time, so such fluctuations are subdominant to experimental uncertainty and can be neglected. 
Fundamental parameters then exhibit a spatial dependence $\langle X(\Vec{x},t) \rangle$  inherited from the time-averaged DM profile as
\bea
\langle X(\Vec{x},t) \rangle = X_0 \l 1 + d_X^{(2)} \frac{\langle \phi^2(\vec{x},t) \rangle}{2} \r. 
\eea
And while fundamental parameters do exhibit oscillations on a time-scale associated with the dark matter mass, in the limit we consider, these fluctuations wash out. Instead, the spatial dependence of these parameters can be probed as experiments traverse an inhomogeneous dark matter background.

\subsection{Matter effect}
\label{Sec: matter effect}
There has been recent interest in effects related to the scattering of quadratically coupled dark matter from macroscopic bodies. Efforts have been made to understand detection prospects for the force induced by the dark matter wind \cite{Fukuda:2018omk, Fukuda:2021drn,Day:2023mkb} as well as the wake force that results from the scattering of DM waves from macroscopic bodies \cite{VanTilburg:2024xib,Barbosa:2024pkl,Grossman:2025cov,Burrage:2025grx}. We take a different but related route.  Instead of characterizing the (wake) force (field gradients), our focus is instead on the field values themselves around macroscopic objects, which induce variations in fundamental couplings.

 Several recent works have characterized the general behavior of scalar fields with interactions that decrease or increase their effective mass near Earth \cite{Banerjee:2025dlo, delCastillo:2025rbr}. We extend these analyses by providing explicit angular and altitude-dependent profiles for the field that result from DM scattering off the Earth and by discussing the implications for searches that probe variations of fundamental constants.   We will be primarily interested in the field values at altitudes corresponding to space station/satellite orbits.

\subsubsection{Profile around Earth}
\label{sec:profile}
Since ultralight dark matter can be treated as a classical field, we can model the scattering problem by solving the Klein-Gordon equation 
\bea \label{eq: KG}
(\Box + m_{\rm DM}^2(\Vec{x})) \phi (\Vec{x},t) = 0
\eea
where the dark matter mass, incorporating matter effects, is given by 
\bea
m_{\rm DM}^2(\Vec{x}) = m_{\rm DM}^2 + \sum_{X} d_{X}^{(2)} Q_{X}(\Vec{x}) \kappa^2 \rho_{\rm SM}(\Vec{x}) \equiv   m_{\rm DM}^2 + \Delta m^{2}_{DM}(\Vec{x}).
\eea
Here, 
\bea \label{eq: dilatonic charge}
Q_{X}(\Vec{x}) = \sum_{A,Z} \frac{\rho (\Vec{x}, A,Z)}{\rho_{\rm SM} (\Vec{x})} \frac{X}{m_{\rm atom} (A,Z)} \frac{\partial m_{\rm atom} (A,Z)}{ \partial X}
\eea
is a dilatonic charge and depends on the matter composition at a given location \cite{Damour:2010rm, Damour:2010rp}. The sum over atomic ($Z$) and mass ($A$) number accounts for each participating isotope. The ratio $\rho (\Vec{x}, A,Z)/\rho_{\rm SM} (\Vec{x})$ determines the fractional abundance of a given isotope specified by $(A,Z)$; $X$ labels the different couplings that appear in Eq. \eqref{dilaton_couplings}; $m_{\rm atom}$ is the mass of a given isotope, and $\rho_{SM}$ is the energy density of the Standard Model matter. The problem of finding the field is simplified if we take the Earth to be a sphere with uniform density.\footnote{The space-borne experiments we consider are performed at altitudes $\gtrsim 400 \rm km$ and away from geographic poles. This is  sufficiently high to average over Earth's inhomogeneities, and for the parameter space in which we are interested, we are insensitive to the interior composition.}
\bea 
(\Box + m_{\rm DM}^2) \phi (\Vec{x},t) = 0  \qquad r > R_\oplus \\
(\Box + m_{\rm DM, \oplus}^2) \phi (\Vec{x},t) = 0 \qquad r \leq R_\oplus. 
\eea
The solution at distances $r\gg R_\oplus$ reduces to a superposition of plane waves that follow the virialized DM momentum distribution. The full solution is therefore  given by
\bea
 \phi(\Vec{x},t) = \int d^3k \sqrt{f(\Vec{k})} \Tilde{\phi}(\Vec{x},t;\Vec{k}), 
 \eea
where $\Tilde{\phi}(\Vec{x},t;\Vec{k})$ is a solution of Eq.~\eqref{eq: KG} given an incoming plane-wave of momentum $\Vec{k}$, and $f(\Vec{k})$ is the boosted Maxwell-Boltzmann distribution
\bea
\label{eq:BoostedMB}
f(\Vec{k}) = \l \frac{1}{2 \pi \sigma_k^2} \r^{3/2} \exp\l -\frac{(\Vec{k}-\Vec{k_0})^2}{2 \sigma_k^2} \r.
\eea
Here, $\Vec{k_0}$ is the average momentum of the DM wind and $\sigma_{k}$ is the velocity dispersion, which throughout the work we approximate as $\sigma_{k}\approx k_0/\sqrt{2}$.

To analyze the scattering problem, we first factorize the field into time- and space-dependent parts:
\bea
\Tilde{\phi}(\Vec{x},t;\Vec{k}) \equiv \text{Re}(e^{i \omega t +\chi_{\Vec{k}}} \psi(\Vec{x};\Vec{k})),
\eea
where $\chi_{\Vec{k}}$ is a random phase, which is independently drawn for each  momentum $\Vec{k}$. 
Using $\omega^2 \equiv k^2 + m^2_{\rm DM}$, the equation of motion for $\psi$ is given by
\bea 
(\nabla^2 + k^2) \psi (\Vec{x};\Vec{k}) = 0  \qquad r > R_\oplus, \\
(\nabla^2 + k^2 - \Delta m_{\rm DM, \oplus}^2) \psi (\Vec{x};\Vec{k}) = 0 \qquad r \leq R_\oplus. 
\eea
We perform a partial-wave expansion to analyze the scattering for each $\psi (\Vec{x};\Vec{k})$.  That is, we decompose each incoming plane wave into states of definite angular momentum and find the coefficients of the scattered waves by matching the field and its derivative at the Earth-vacuum boundary. This is the standard quantum mechanics problem of scattering from a spherical potential.  We write: 
\bea
\label{eq:multipole}
\psi = \vert \psi_0 \vert \sum_{l}(2l+1)i^l  P_l \l \cos \theta \r \left\{
\begin{matrix}
 j_l(kr) + A_l h_l(kr), & r\geq R_\oplus \\
B_l j_l(k'r), & r<R_\oplus
\end{matrix} ,\right.
\eea
where $\psi_0$ is the amplitude of the incoming plane-wave, $\theta$ is the angle between $\vec{k}$ and $\vec{x}$, and $k' \equiv \sqrt{k^2 - \Delta m_{\rm DM, \oplus}^2}$. Here, $j_{l}$ is the spherical Bessel function and $h_{l}$ is a spherical Hankel function. In spherical coordinates, $\cos \theta$ is
\bea
\label{eq:dotproduct}
\cos \theta = \uvec{k} \cdot \uvec{x} = \sin \theta_k \cos \varphi_k \sin \theta_r + \cos \theta_k \cos \theta_r.
\eea
Here, we have defined $\theta_{k}$  ($\theta_{r}$) as the angle between $\Vec{k}$ ($\Vec{x}$) and $\Vec{k_0}$, and $\varphi_k$ is the azimuthal angle of $\Vec{k}$, while for $\Vec{x}$ we set $\varphi_r = 0$ due to the azimuthal symmetry of the problem.
 The coefficients $A_l$ and $B_l$ are given by
\bea \label{eq: A, B coefficients}
A_l = \frac{j_l'(kR_\oplus) j_l(k'R_\oplus) - j_l(kR_\oplus) j_l'(k'R_\oplus)}{h_l(kR_\oplus) j_l'(k'R_\oplus) - h_l'(kR_\oplus) j_l(k'R_\oplus)},
\, \, \,
B_l = \frac{h_l(kR_\oplus) j_l'(kR_\oplus) - h_l'(kR_\oplus) j_l(kR_\oplus)}{h_l(kR_\oplus) j_l'(k'R_\oplus) - h_l'(kR_\oplus) j_l(k'R_\oplus)}.
\eea
Here, $j_l'(z)$ and $h_l'(z)$ are derivatives of the corresponding Bessel and Hankel functions with respect to the argument $z$. To find the expectation value of the field around Earth, we average over time and phases $\chi_{\Vec{k}}$:
\bea
\label{eq:averagedphi}
\phi^2(\Vec{x}) \equiv \langle \phi^2(\Vec{x},t) \rangle &=& \int d^3p \, d^3k \sqrt{f(p)f(k)} \langle  \Tilde{\phi}(\Vec{x},t;\Vec{p}) \Tilde{\phi}(\Vec{x},t;\Vec{k})\rangle \\
&=& \int d^3k f(k) \vert \psi(\Vec{x},\Vec{k}) \vert^2 .
\eea
In the last line, we have used the fact that terms with $\vec{k} \neq \vec{p}$ average to zero, owing to the fact that they would have independent phases, i.e. the averaging over phases effectively contributes a $\delta^{(3)}(\vec{k} - \vec{p})$.
Before presenting numerical results, we now turn to a semi-analytic discussion of the  large coupling limit. 

\subsubsection{Profile at large coupling}
Our interest in this paper will be in the regime $k_0^2 \ll \Delta m_{\rm DM, atm}^2 \ll \Delta m_{\rm DM, \oplus}^2$. 
In this case, ground-based experiments suffer from suppressed $\phi$ field values, and so space-based experiments are particularly useful. The field essentially reflects from the Earth's surface/atmosphere, and the problem reduces to that of scattering off of a hard sphere.\footnote{ The precise height where this reflection occurs depends on the strength of the coupling and the DM mass.  The atmospheric density changes as $\sim \exp (- h/8\,\rm km)$; this can shift the scattering surface by as much as $150\, \rm km$ for the most extreme coupling-mass values we consider. This is smaller than the lowest practical orbit of $\sim 400 \, \rm km$. For simplicity, we phrase the problem as reflection from the Earth's surface. Modelling this effect in more detail would lead to a modest increase in sensitivity beyond the results shown here.} 
  
We can impose a Dirichlet boundary condition, which yields 
\bea
A_l = -\frac{ j_l(kR_\oplus) }{h_l(kR_\oplus) }, \quad
B_l = 0.
\eea
This step is equivalent to taking the limit $k' R_\oplus \to i \infty$ in Eq. \eqref{eq: A, B coefficients}.
We will take advantage of this approximation to understand the field profile in the low momentum ($kR_\oplus \ll 1$, $kr\ll1$) limit near the Earth's surface. 
For small arguments, the Bessel functions scale as
\bea
j_l \propto (kr)^l, \quad h_l \propto (kr)^{-l-1},
\eea
so terms of lowest $l$ contribute the most.   In this limit, the leading $A_l$ simplify to
\bea
A_0 \approx i kR_\oplus + (kR_\oplus)^2, \quad A_1 \approx i \frac{(kR_\oplus)^3}{3}, ...
\eea
While $l=0$ dominates, we keep the $l=1$ term. It is of potential experimental significance: it encodes the leading directional dependence. Keeping the leading terms in $kR_\oplus$ and retaining only the $l=0,1$ contributions in the partial wave expansion, the spatial dependence of the field profile is given by
\bea
\frac{\vert \psi(\Vec{x},\Vec{k}) \vert^2}{\vert \psi_0 \vert^2} \approx 
  \l 1-\frac{R_\oplus}{r}\r^2 P_0(\cos \theta) + 2 (k R_\oplus)^2\l 1-\frac{R_\oplus}{r}\r\l\frac{r}{R_\oplus}-\l \frac{R_\oplus}{r} \r^2\r P_1 (\cos \theta),
\eea
where $P_{l}$ are the Legendre polynomials.  This expression gives the field profile for a monochromatic incoming wave. The expression for the profile of interest is found by convolving this monochromatic result with the boosted Maxwell-Boltzmann distribution, Eq.~(\ref{eq:BoostedMB}), taking care to average over the relevant angles $\phi_{k}$ and $\theta_{k}$, see Eq.~(\ref{eq:dotproduct}). This yields:
\bea
\phi^2(\Vec{x}) \approx \frac{\rho_{\rm DM}}{m_{\rm DM}^2}  \l \l 1-\frac{R_\oplus}{r}\r^2 +  \frac{C}{4}(k_0 R_\oplus)^2   \l 1-\frac{R_\oplus}{r}\r\l\frac{r}{R_\oplus}-\l \frac{R_\oplus}{r} \r^2\r P_1 (\cos \theta_r) \r \\ + \mathcal{O}((k_0 R_\oplus)^2), \nonumber
\eea
where $C = \frac{9}{e\sqrt{\pi}}+\frac{21\, \text{Erf}(1)}{2}$. We write $r = R_\oplus +h$,  and close to the surface ($h\ll R_\oplus$) we can expand in $h/R_{\oplus}$.  The result is:
\bea
\label{eq:phiSqApprox}
\phi^2(\Vec{x}) \approx \frac{\rho_{\rm DM}}{m_{\rm DM}^2} \l \frac{h}{R_\oplus} \r^2 \l 1 +  C(k_0 R_\oplus)^2  P_1 (\cos \theta_r) \r + \mathcal{O}((k_0 R_\oplus)^2).
\eea
The spherically symmetric (i.e., $P_{0}$) term is well known, while the second term is new. It is unsurprising that the $P_{0}$ term dominates in this ($R_\oplus \ll 2\pi k_0^{-1}$) limit, as the Earth is effectively point-like when compared to the de Broglie wavelength of the DM.
Note, the $P_1$ term is effectively independent of the mass, so the angular dependence of the field is also approximately mass-independent. A straight-forward consequence is that experiments that rely on this $P_1$ term to limit $\phi$-SM couplings produce mass-independent limits in the low-mass regime where Eq.~(\ref{eq:phiSqApprox}) applies.

For larger masses, $k_0 R_\oplus \sim  1$, the above approximation breaks down, and the profile must be evaluated numerically. It is still useful to expand the field in terms of Legendre polynomials as
\bea \label{eq: phi^2 legendre}
\phi^2(\Vec{x})  = \frac{\phi_{0}^2}{2} \sum_{l=0}^{N} a_l (h,m_{\rm DM}) P_l(\cos \theta_r).
\eea
Now, the coefficients $a_l(h,m_{\rm DM})$ are obtained via a numerical calculation, whose procedure is outlined in Appendix \ref{App: multipole expansion}. They depend on the height $h$ above the Earth's surface and the mass $m_{\rm DM}$ of the dark matter.  In practice, we find $N=4$ sufficient to characterize the field to $1\%$ precision at any height. 

\begin{figure}
    \centering  \includegraphics[width=\columnwidth]{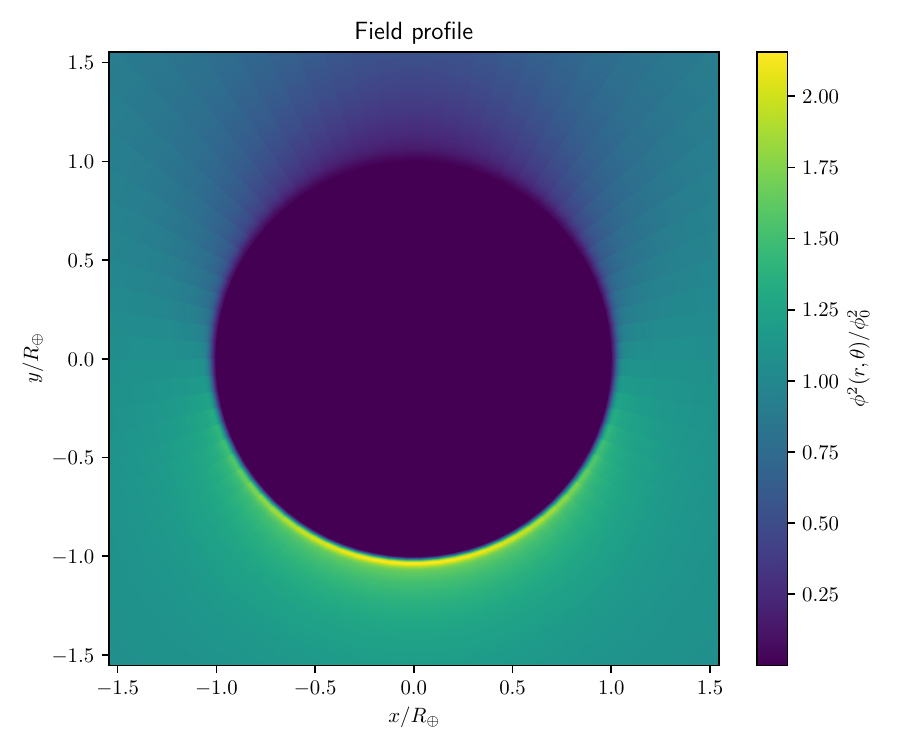}
    \caption{Example field profile due to scattering of dark matter from the Earth's surface for dark matter mass $m_{\rm DM} \approx 10^{-9} \, \rm eV$ and coupling $d_e \approx 2 \times 10^{17}$ (or $d_{m_e} \approx 1 \times 10^{18}$ or $d_{g} \approx 4 \times 10^{14}$ or $d_{\hat{m}} \approx 3 \times 10^{16}$).  The $x-$ and $y-$ axes are plotted in units of Earth radii $R_{\oplus}$. The color bar indicates the square of the field value $\phi^2(r,\theta)$ in units of the (field value)$^2$ at infinity  $\phi_0^2$. The dark matter wind is incident from the bottom of the figure; the induced shadow is  visible above the Earth.}
    \label{fig: field profile}
\end{figure}

\begin{figure}
    \centering
\includegraphics[width=0.88\columnwidth]{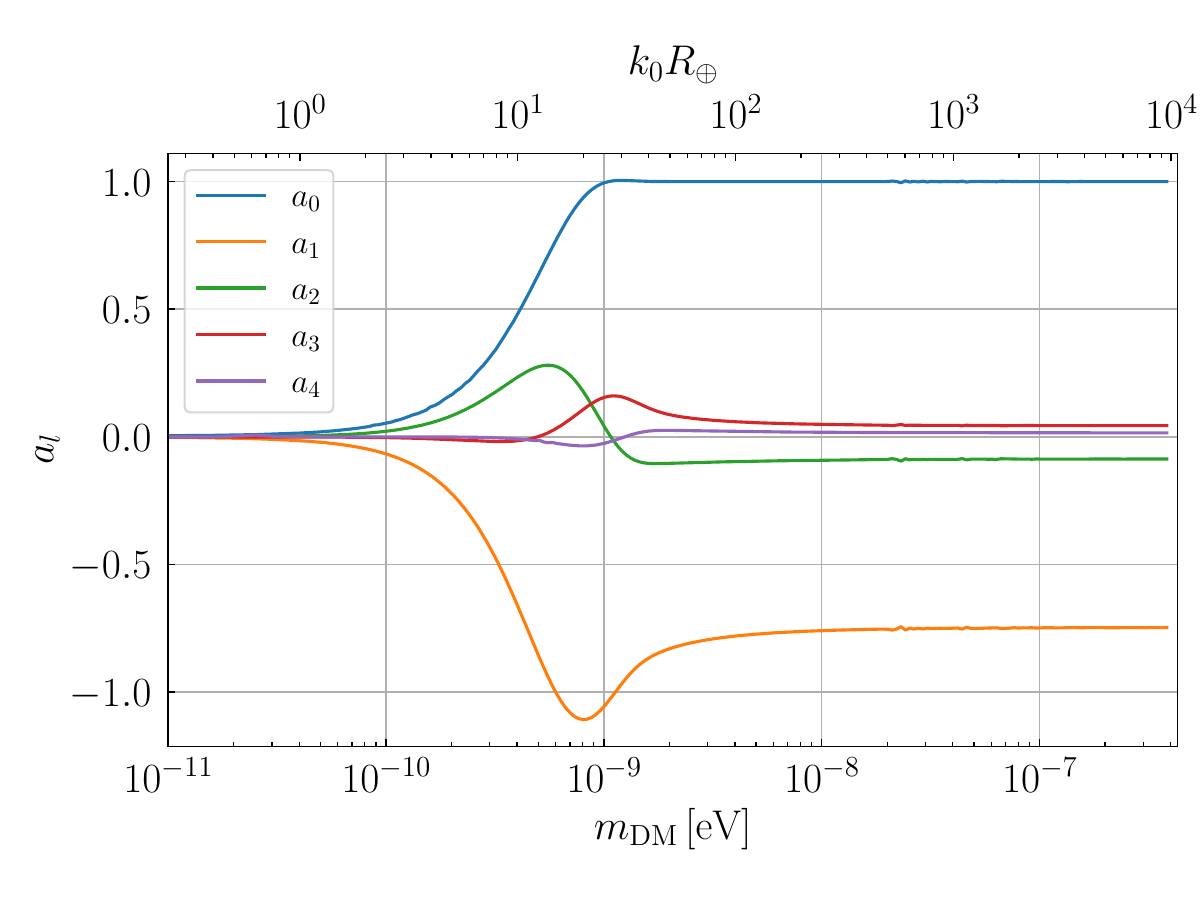}
    \caption{Legendre coefficients $a_l(h,m_{\rm DM})$ for the field profile (see Eq.~(\ref{eq: phi^2 legendre})) at $h = 0.06\, R_\oplus$, the height of the ISS orbit, as a function of the dark matter mass $m_{\rm DM}$. The upper axis gives the corresponding value of the average momentum of the DM wind $k_0$ in units of inverse Earth radii. }
    \label{fig: Legendre}
\end{figure}

An example field profile is shown in Fig. \ref{fig: field profile}. We show the square of the field value $\phi^2(r,\theta)$, see Eq.~(\ref{eq:averagedphi}), in units of the field value at infinity squared $\phi_0^2$.  The $x-$ and $y-$axes are plotted in units of Earth radii.  An enhancement in the DM density can be seen (due to interference) on the side where the dark matter wind is incident (bottom of the Figure), while the shadow is present at top.  Here, we have chosen $k_0 R_\oplus = 30$, $\Delta m_{\rm DM,\oplus}^2/k_0^2 = 900$, which is equivalent to roughly $m_{\rm DM} \approx 10^{-9} \, \rm eV$ and $d_e \approx 2 \times 10^{17}$ (or $d_{m_e} \approx 1 \times 10^{18}$ or $d_{g} \approx 4 \times 10^{14}$ or $d_{\hat{m}} \approx 3 \times 10^{16}$).
In Fig.~\ref{fig: Legendre} we show how these $a_l(h,m_{\rm DM})$ vary with the mass of the field for a fixed $h=0.06 R_{\oplus}$, corresponding to the height of the International Space Station (ISS) orbit, which provides a (potential) home for some of the atomic clock-based measurements discussed in Sec.~\ref{sec:Space To Ground} and Sec.~\ref{sec:Space to Space}.  At very large values of $kR_{\oplus}$, the geometric-optics limit applies, and so the $a_{l}$ become independent of the precise value of the mass.  This behavior can be seen at the largest value of masses in Fig.~\ref{fig: Legendre} for fixed $h$. Table \ref{tab: legendre coefficients} provides asymptotic values ($m_{\rm DM} \gg 10^{-8} \text{ eV}$) of the $a_l(h,m_{\rm DM})$ coefficients at two heights, $h_{\rm ISS} = 0.06 R_\oplus$  and $h = 0.1 R_\oplus$. The value $h=0.1 R_\oplus$ corresponds to the MICROSCOPE satellite \cite{Touboul:2017grn}, which gives a sensitive test of the weak equivalence principle that can be reinterpreted as a bound on $\phi$ \cite{Hees:2018fpg,Banerjee:2022sqg}. These coefficients may be used to evaluate the field profile for masses $m_{\rm DM} \gg 10^{-8} \, \rm eV$, where a full solution to Eq.~(\ref{eq: KG}) is more computationally intensive. 

\begin{figure}
    \centering
 \includegraphics[width=0.91\columnwidth]{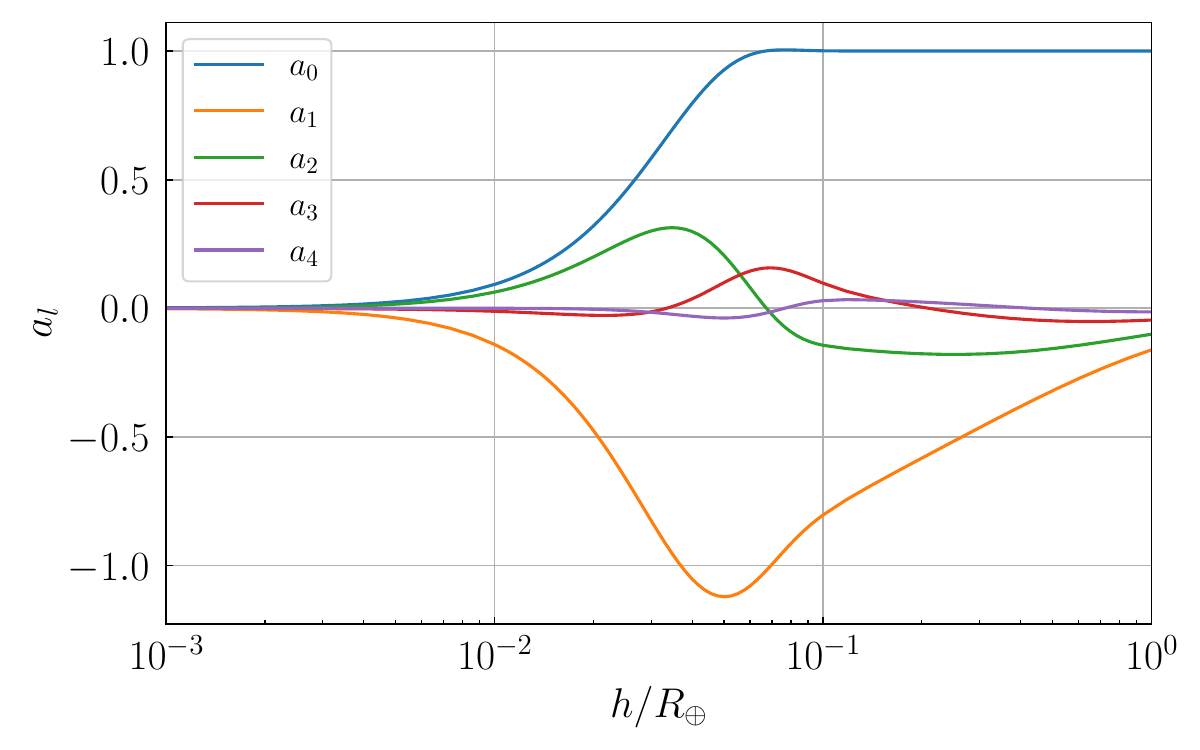}
    \caption{Legendre coefficients $a_l(h,m_{\rm DM})$ for the field profile (see Eq.~(\ref{eq: phi^2 legendre})) evaluated at $m_{\rm DM} = 10^{-9}\, \rm eV $ as a function of height $h$ measured in Earth radii $R_\oplus$. The ISS orbit is at $h= 0.06 R_\oplus$, while the MICROSCOPE experiment orbit corresponds to $h=0.1 R_\oplus$.}
    \label{fig: Legendre height}
\end{figure}

In Fig. \ref{fig: Legendre height}, we show how the $a_l$ coefficients depend on the altitude $h$ in units of Earth radii $R_{\oplus}$. The most important terms are the monopole ($a_0$) and the dipole ($a_1$). When $k_0 h \ll 1$, their magnitudes grow as $h^2$. When $k_0 h \gtrsim 1$, the monopole saturates at $a_0 = 1$, so experiments looking for angle-averaged offsets between the field value in space and on Earth will not  benefit by going to higher altitudes. However, the same is not true for the dipole term $a_{1}$.  It is most pronounced when $k_0 h \sim 1$ but then it begins to fall off, and at $h\sim R_\oplus$ it is already suppressed by an order of magnitude. Thus, to take advantage of the dipole feature experiments should be performed at altitudes $h\lesssim 1000 \, \rm km$, where the influence of the Earth is still large. In the next section we discuss in some detail how the dipole term modulates potential signals.
\begin{table}[t] 
\centering
 \begin{tabular}{|c | c c c c c|} 
 \hline
  $h/R_\oplus$ & $a_0$ & $a_1$ & $a_2$ & $a_3$ & $a_4$ \\ [0.5ex] 
 \hline
  0.06  & $1$ & $-0.75$ & $-0.09$ & $0.04$ & $0.02$ \\ 
 0.1 & $1$ & $-0.67$ &  $-0.11$ & $0.02$ & $0.02$ \\ %[1ex] 
 \hline
 \end{tabular}
 \caption{Legendre coefficients for the field profile (see Eq.~(\ref{eq: phi^2 legendre})), valid for $m_{DM} \gg 10^{-8}$ eV.}\label{tab: legendre coefficients} 
\end{table}

\section{Probing dark matter with clocks}\label{sec:clocks}

Having discussed the inhomogeneous profile of the dark matter field, we are in a position to analyze specific experimental set-ups and their sensitivity to dark matter.

One way to look for variation of fundamental parameters is to monitor energy levels in quantum clocks \cite{Arvanitaki:2014faa,Antypas:2022asj}. The sensitivity of a clock depends on how sensitive the clock's transition frequency $f$ is to the dark matter-induced change in an underlying fundamental parameter $X$,
\bea \label{eq: parameter to freq}
\frac{\Delta f}{f} = K_X \frac{\Delta X}{X}.
\eea
Here, $K_X$ is an enhancement factor and $X$ is a fundamental parameter on which the frequency might depend, e.g. the fine-structure constant $\alpha$ or electron mass $m_{e}$. The correction to the value of a fundamental parameter $X$ is related to the background density of the DM through (see Eq.~(\ref{eq: phi^2 legendre}))
\bea \label{eq: dm to parameter}
\left\langle \frac{\Delta X}{X} \right\rangle = \frac{d^{(2)}_X \kappa^2 \phi^2(\Vec{x})}{2} =    \frac{d^{(2)}_X \kappa^2 \phi_{0}^2}{4} \sum_{l=0}^{N} a_l (h,m_{\rm DM}) P_l(\cos \theta_r). 
\eea

Observing such a frequency shift requires a reference frequency against which to compare. This can be provided, for instance, by a copy of the clock located where the signal is absent. For our parameters of interest, a clock close to the surface of the Earth would serve this purpose.  A (space-borne) clock exposed to the dark matter will see different values of the fundamental parameters.  So, despite their identical architecture, the clocks will tick at different frequencies.  The dark matter can then be searched for either by comparing directly how the frequencies $f_1$ and $f_2$ of two clocks are affected, or by measuring how their relative phases, $\Delta \varphi = 2\pi \int (f_1(t) - f_2(t)) dt$, drift. This setup will be the topic of Sec.~\ref{sec:Space To Ground}. 
An alternative experimental configuration involves two different clocks placed in the same DM background.  When the two clocks utilize different transitions for their reference frequency, their responses to variations in the local dark matter density differ, i.e. the relevant transition energies depend on the dark matter profile in different ways.  This approach is the subject of Sec.~\ref{sec:Space to Space}.

\subsection{Sampling of $\cos \theta_r$} \label{sec: cos thetar}
Before turning to the details of particular experimental set-ups, it is useful to understand exactly how clocks sample the inhomogeneous dark matter profile. Recall, we have defined $\theta_{r}$ as the angle between $\vec{k_0}$ (the average DM wind momentum) and the position $\vec{x}$.  In Sec.~\ref{sec:background}, we gained insight into the field profile as a function of $\cos \theta_r$, and showed how in the high-mass limit the field exhibits a pronounced dipole feature. Orbital experiments can take advantage of this feature as they scan values of $\cos \theta_r$ as a function of time.\footnote{The $\cos \theta_r$ dependence of the signal can be mapped onto a time-dependent signal as the position of the space-clock is known to a high precision at any instant. When $m_{\rm DM} \gg 10^{-9} \, \rm eV$, the coherence time of the DM is much shorter than the duration of the Ramsey sequence, $\mathcal{O}(1 \, \rm  s) $.  In this case, fluctuations due to the finite coherence time of the dark matter field  average out and a frequency measurement at a given location essentially corresponds to a known  DM amplitude. For smaller masses, $10^{-12} \, \rm eV \lesssim m_{\rm DM} \lesssim 10^{-9} \, \rm eV$, the value of the DM signal may experience fluctuations around the expectation value. In this case, the expectation value, and hence the apparent coherent time-dependence, only emerges after averaging over $T_{\rm int} \gg t_{\rm coh}$. Since this condition is satisfied for all experiments we discuss, we use spatial- and temporal-dependence interchangeably.  }   
This discussion is of interest for experiments where direct comparisons of frequency are made. We discuss phase measurement in Sec.~\ref{sec:phase}, where the detailed $\cos \theta_r(t)$ dependence will not be relevant. 

\begin{figure}
    \centering
    \includegraphics[width=0.6\columnwidth]{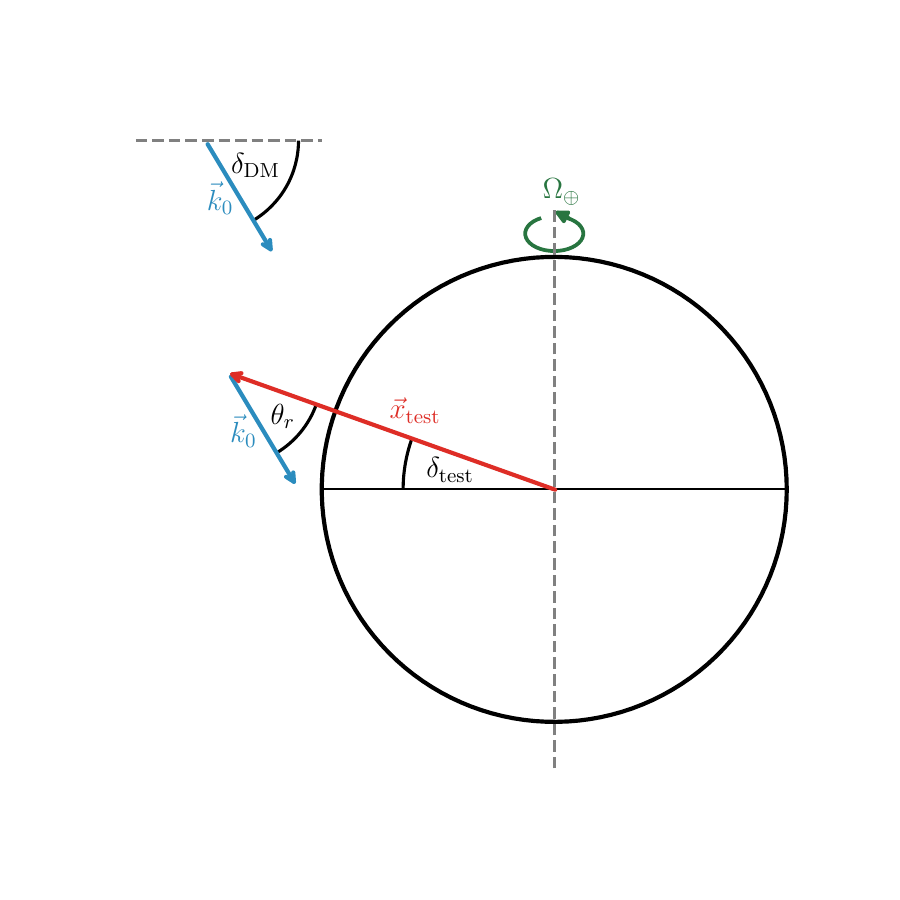}
    \caption{Schematic representation of coordinates in the Earth-centered inertial frame at an instant when  $\alpha_{\rm test}(t) = \alpha_{\rm DM} + \pi$.  The dark matter wind is shown as incident from the upper-left with momentum vector $\Vec{k_0}$.  The Earth rotates about its axis with frequency $\Omega_{\oplus}$. The declination of the incident wind and the experiment are given by $\delta_{\rm DM}$ and $\delta_{\rm test}$, respectively. }
    \label{fig: Schematic}
\end{figure}

 The way in which experiments sample $\cos \theta_r$ need not be just a function of orbital parameters, but can also depend on the details of the experimental protocol employed.  In particular, it will
 depend upon whether an experiment compares frequencies of two co-located clocks in space or whether one clock is in space and the other is on Earth. We will study experiments performed in LEO with low eccentricity. 
%%%%
Throughout the rest of this work we  operate in the Earth-centered inertial frame, which is the common choice for space-based experiments.   Fig. \ref{fig: Schematic} provides a schematic representation of the coordinates in this frame.  

%%%
%%%%

The direction of the DM wind (shown at upper left) is approximately constant. (We ignore  seasonal modulation of the wind direction; the dominant time dependence comes from the motion of the experiment. The annual modulation from the Earth's orbital velocity changes the effective wind direction and speed at the $\mathcal{O}(v_\oplus/v_{\rm DM})\sim 10\%$ level and can be incorporated into a time-dependent signal template.) The Earth rotates with frequency $\Omega_{\oplus}$. In this frame, the coordinates of the test body position $\Vec{x}_{\rm test}$ and the average DM wind momentum are given by

\bea
\Vec{x}_{\rm test}(t) = 
\begin{pmatrix}
r_{\rm test} \cos \delta_{\rm test}(t) \cos \alpha_{\rm test}(t) \\
r_{\rm test} \cos \delta_{\rm test}(t) \sin \alpha_{\rm test}(t) \\ 
r_{\rm test} \sin \delta_{\rm test}(t) 
\end{pmatrix}, \qquad
\Vec{k_{0}} = 
\begin{pmatrix}
k_{0} \cos \delta_{\rm DM} \cos \alpha_{\rm DM} \\
k_{0} \cos \delta_{\rm DM} \sin \alpha_{\rm DM} \\ 
k_{0} \sin \delta_{\rm DM} 
\end{pmatrix}.
\eea
Here $\delta_x$ is a declination angle, and $\alpha_{x}$ is the right ascension measured with respect to the Earth. For concreteness, we take direction of the DM wind to be inverse of the direction towards the Cygnus constellation \cite{Bandyopadhyay:2010zj}, which yields $\delta_{\rm DM} = - 42^\circ$, $\alpha_{\rm DM} = 51^\circ$.
We can write the time-dependent value of $\cos \theta_r$ in terms of declination and right ascension as: 
\bea
\cos \theta_r(t) = \Vec{\hat{x}}_{\rm test} \cdot \Vec{\hat{k}}_{0} = -\cos \l \delta_{\rm DM} + \delta_{\rm test} (t) \r \l\frac{1- \cos \l \alpha_{\rm DM} - \alpha_{\rm test}(t) \r}{2}\r \nonumber \\
+\cos \l \delta_{\rm DM} - \delta_{\rm test} (t) \r \l\frac{1+ \cos \l \alpha_{\rm DM} - \alpha_{\rm test}(t) \r}{2}\r.
\eea

\begin{figure}
    \centering
\includegraphics[width=0.7\columnwidth]{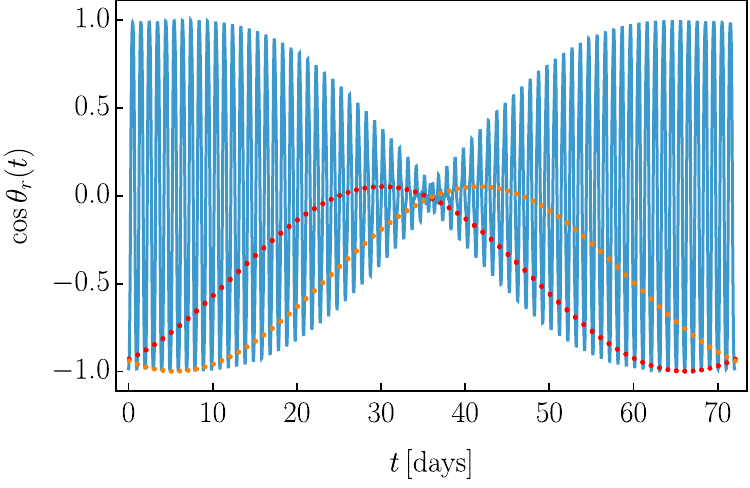}
    \caption{Example time-dependence of $\cos \theta_r$, see Eq.~(\ref{eq: costheta}), with $\delta_0 = 52 ^{\circ}$.
    and $\omega/\Omega=72$. For the case of a space-ground comparison, the red (orange) dots represent the $\cos \theta_r$ values sampled over the course of a single day when the satellite is descending (ascending). The Earth-based clock is located at $45^{\circ}$ N.}
    \label{fig: cos theta}
\end{figure}

For co-located clock systems, measurements are taken at equal time intervals, which means that the modulation of $\cos \theta_r(t)$ by the orbit is imprinted directly on the signal measured by clocks. The evolution of the declination and right ascension of the clock can be parametrized as

\bea
\delta_{\rm test} (t) &=& \delta_0 \cos{\omega t} \\
\alpha_{\rm test} (t) &=& (\Omega  + \omega) t + \alpha_{\rm DM}  \qquad \rm{(mod \; 2\pi)}.
\eea
Here, $\omega$ is the angular frequency of the orbit, and $\Omega$ is defined as the angular frequency of precession of the orbit.\footnote{Due to the Earth's shape, all orbits that do not cross through the pole experience this precession.} So, 
\bea \label{eq: costheta}
\cos \theta_r(t) = 
\sin  \delta_{\rm DM}  \sin \l  \delta_0 \cos  \omega t   \r 
+ \cos  \delta_{\rm DM}   \cos \l \delta_0 \cos  \omega t   \r  \cos \l (\Omega  + \omega) t \r .
\eea
Fig. \ref{fig: cos theta} shows an example of  $\cos \theta_r(t)$ for a space-borne experiment as a blue curve.  We use $\delta_0 = 51.6^{\circ}$ which corresponds to the orbit for the ISS. The modulation of the amplitude is due to precession of the orbital axis around Earth's rotation axis. This precession results in an oscillation between the angle between the DM wind ($\delta_{DM} =-42^{\circ}$) and the  orbital axis.  To (over-)emphasize this effect, we have plotted  $\cos \theta_{r}(t)$ using a reduced ratio $\omega/\Omega$ = 72; this is roughly an order of magnitude smaller than the true ISS value $(\omega/\Omega)\vert_{\rm ISS} \approx 1200$ (throughout this work we assume $\omega_{\rm ISS} \approx 16 \times 2\pi \,  \text{day}^{-1}$ and $\Omega_{\rm ISS} \approx 2\pi /72 \, \text{day}^{-1}$). This choice allows us to more clearly display how modulation of the amplitude proceeds. For the particular case of the ISS orbit, because $\delta_{0}-\delta_{DM}\simeq 90^{\circ}$ and $\delta_{0}+ \delta_{DM} \simeq 0$, the DM wind can transition from being essentially parallel to the orbit to orthogonal to the direction of the orbit during a single precession period. When these vectors are perpendicular,  $|\cos \theta_r(t)|$ is maximal, while for an anti-aligned orientation, all points on the orbit experience $\theta_r \approx \pi/2$, and $|\cos \theta_r(t)|$  is suppressed during such an orbit.  This is seen in Fig.~\ref{fig: cos theta} at $t\approx 36$.  The modulation of the amplitude has implications for time-varying signal searches that  will be discussed in the following sections.

When frequency comparisons are made between space- and Earth-based clocks, they must be in visual contact. So, $\cos \theta_{r}(t)$ is preferentially sampled at these times.  This occurs
when their coordinates are approximately equal:
\bea
\delta_{\rm test} (t) &=&  \delta_0 \cos  \omega t  \approx \delta_{\rm lab},\\
\alpha_{\rm test} (t) &=& (\Omega  + \omega) t + \alpha_{\rm DM}\approx \alpha_{\rm lab} (t) = \Omega_\oplus   t + \alpha_{\rm DM} \qquad \rm{(mod \; 2\pi)}.
\eea
The first condition is satisfied when 
\bea
\omega t_{\pm,n} = \pm \beta + 2\pi n
\eea
when $n$ is the number of the orbit and $\pm$ determines whether the satellite is descending ($+$) or ascending ($-$) with respect to the equator; this expression defines the phase $\beta$. The second condition is not satisfied for every $n$  (i.e. for some orbits the ascension does not line-up at the moment the declination does), but any time that the two clocks are in contact the satellite is in one of two phases: $\beta$ or $-\beta$. If the clock's orbital precession period is much longer than the duration of the experiment, such a setup effectively probes the field value at the two locations.\footnote{When $\beta$ is close to $0$ or $\pi$ the phases cannot be resolved and the satellite probes only a single location.} However, as orbital precession proceeds, space-to-ground comparisons experience a modulation of the signal at the frequency of orbital precession
\bea \label{eq: costheta 2}
\cos \theta_r(t_{\pm,n}) = 
\sin  \delta_{\rm DM}  \sin   \delta_{\rm lab}   
+ \cos  \delta_{\rm DM}   \cos  \delta_{\rm lab}    \cos \l \frac{\Omega}{\omega} 2\pi n  \pm \alpha_0\r 
\eea
where $\pm\alpha_0$ is the right ascension at $t_{\pm,0}$, and $\alpha_0 = \beta (1+ \Omega/\omega)$. Since the period of orbital precession is typically much longer than the period of a single orbit, $\Omega/\omega \ll 1$, we can make the approximation $\alpha_0 \approx \beta$. 
In Fig.~\ref{fig: cos theta} we show a simplified picture illustrating how the sampling works for a laboratory located at $45^{\circ}$ N. In the figure, the red (orange) dots represent values of $\cos \theta_r$ sampled over the course of a single day when the satellite is descending (ascending), i.e. $\pm \beta$.  In practice, the actual number of measurements is larger than that suggested by the figure, since each clock pair can perform roughly 5-6 space-to-ground comparisons throughout the day, roughly 3 times per descending/ascending phase \cite{Savalle:2019isy}. These occur when the coordinates are ``close enough" that contact is maintained.   

The above expression (and Fig.~\ref{fig: cos theta}) implies that these space-to-ground comparisons do not sample the full range of values of  $\cos \theta_r$.  Moreover, the range of these values depends on the location of the Earth-based clock. Therefore, when performing frequency comparisons between space- and Earth-based clocks, placement of the latter can impact the DM signal and should be taken into account when optimizing for static/time-varying signals. 

\subsection{Space-to-ground experiments} 
\label{sec:Space To Ground}
The first setup we discuss involves one clock in orbit around the Earth, either onboard the ISS \cite{Savalle:2019isy, Schkolnik:2022utn} or as an independent satellite \cite{Derevianko:2021kye}. The second clock is located on Earth.  Such a setup allows for multiple tests of fundamental physics, including deviations from gravity and dark matter searches \cite{Delva:2018ilu,Derevianko:2021kye}. 

In our discussions, we make the self-consistent assumption that the field is negligible close to the Earth's surface. Relative to the ground clock, the frequency of the clock in space is modified not only by the Doppler effect and gravitational redshift, but also by the DM-induced shift (see Eq.~(\ref{eq: dm to parameter})).  

As shown in Fig. \ref{fig: Legendre}, when $m_{\rm DM} \gtrsim 10^{-9} \, \text{eV}$ and the clock is onboard the ISS $(h/R_{\oplus} \simeq 0.06)$, the spherically symmetric component of the field profile ($a_0 = 1$) is the same as in the absence of scattering from macroscopic bodies. For circular orbits this introduces a constant offset between clocks that is independent of the orbital phase. There is an additional strong angular dependence,  $a_1 \sim \mathcal{O}(1)$, which introduces a modulation of the signal on the orbital timescale. 

Depending on the type of clock used, the clock comparison is performed by tracking the relative phase of two clocks \cite{Savalle:2019isy}, a frequency ratio \cite{Schkolnik:2022utn}, or both \cite{Derevianko:2021kye}. Below we describe how the DM signal affects the different types of measurement, and we estimate the sensitivity of each technique.

\subsubsection{Phase measurement}
\label{sec:phase}
We begin by investigating a phase measurement. A phase difference is caused by an underlying frequency mismatch between the clock in space $S$ and one on the ground $G$
\bea
f_{\rm S}(t) - f_{\rm G} = \Delta f_{\rm rel}(t) + \Delta f_{\rm DM}(t).
\eea
The first term on the right-hand side contains the relativistic frequency shifts due to the Doppler effect and gravitational redshift, while the second term is induced by a difference in the dark matter field value in space compared to that on the ground. The expected phase shift due to relativistic effects, $\Delta \varphi_{\rm rel} (t,t_0)$, can be accounted for by a careful characterization of the orbit of the space clock \cite{Savalle:2019isy}. It can subsequently be subtracted from the total phase shift $\Delta \varphi (t,t_0)$.  This allows for the isolation of an anomalous signal, in our case due to dark matter.
The DM-induced phase difference from a time $t_0$ to a time $t_0 + \Delta t$, $\Delta \varphi_{\rm DM} (t_0 + \Delta t,t_0)$, is given as:
\bea
    \Delta \varphi_{\rm DM}(t_0 + \Delta t,t_0) = 2\pi\int\limits_{t_0}^{t_0 + \Delta t}{\Delta f_{\rm DM}(t')\, dt'},
\eea
where the shift in frequency due to the presence of the background DM is given by
\bea \label{Eq: frequency dependence}
\Delta f_{\rm DM}(t') &=&  \Delta f_{\rm DM,0} \sum_l  a_l (h,m_{\rm DM}) P_l \l \cos \theta_r (t) \r \\
&\approx& \Delta f_{\rm DM,0} \l a_0(h,m_{\rm DM}) + a_1 (h,m_{\rm DM}) \cos \theta_r (t) \r. \nonumber
\eea
Here, $\Delta f_{\rm DM,0} \equiv (K_X   d^{(2)}_X \kappa^2 \phi_{0}^2/4)f_0$, and $f_0$ is the frequency of the clock in the absence of DM.
The accumulated phase receives contributions from both the monopole and the dipole term 
\bea
    \Delta \varphi_{\rm DM}(t_0 + \Delta t,t_0) = 2 \pi \Delta f_{\rm DM,0} \Delta t \l a_0(h,m_{\rm DM}) + \frac{a_1(h,m_{\rm DM})}{\omega \Delta t} \Tilde{F}(t_0 + \Delta t,t_0) \r,
\eea
where $\Tilde{F}(t,t_0) = \omega \int_{t_0}^t{\cos \theta_r (t') \, dt'}$ is an oscillatory function with an $\mathcal{O}(1)$ amplitude, $\vert \Tilde{F}(t,t_0) \vert \lesssim 1$. 
Typically, the duration of the experiment is much longer than the orbital period. Then the phase difference between the initial and final time is dominated by the linear phase drift. The oscillatory piece is suppressed as $\omega \Delta t \gg 1$.\footnote{The above expression neglects contributions from $\mathcal{O}(1)$ amplitude fluctuations, see discussion below Eq.~(\ref{eq: simpleosc}).  As noted there, the duration of the experiment becomes long with respect to the coherence time, fluctuations are suppressed, and do not change any of the conclusions presented here.}
This results in an anomalous time drift between clocks $\delta t$ with
\bea \label{eq: time res}
   \frac{\delta t}{\Delta t} =  \frac{\Delta \varphi_{\rm DM}(t,t_0)}{2\pi f_0 \Delta t} =  a_0(h,m_{\rm DM})  \frac{\Delta f_{\rm DM,0}}{f_0}    .
\eea
This signal is identical to the anomalous gravitational redshift that is a primary physics target of the ACES mission. That signal is parametrized as 
\bea
\frac{\delta t}{\Delta t} = \alpha \Delta U,
\eea
where $\alpha$ encodes the deviation from gravity while $\Delta U \approx 3.6 \times 10^{-11}$ is the average gravitational potential difference between the ISS and the clock on Earth. Their science goal is to measure $\alpha$ to $\alpha < 2\times 10^{-6}$, which translates to $\frac{\delta t}{\Delta t} < 7 \times 10^{-17}$.  This can be translated into a sensitivity to the $d_X^{(2)}$ coupling\footnote{To reach such a precision, the ISS location needs to be measured with an accuracy better than $\delta h < 300 \, \rm m$ for a 10 day integration time \cite{refId0,Savalle:2019isy}. This is about 3 orders of magnitude less stringent than the current orbit determination error \cite{montenbruck2011orbit}, and  the requirement scales down with number of orbits as $\delta h \propto 1/N_{\rm orb}$, so even future missions involving $\Delta f / f \sim 10^{-19}$ clocks should not be limited by this systematic. Other environmental and clock-internal shifts, such as thermal, magnetic, collisional, and residual Doppler shifts, enter through the quoted ACES clock and link accuracy budget; the relevant systematic shifts have been studied in detail in Refs.~\cite{Laurent:2015,Laurent2020QualificationAF}. }
via Eqs. \eqref{eq: parameter to freq} and \eqref{eq: dm to parameter}: 

\bea \label{eq: dx phase}
d_{X}^{(2)} \lesssim \frac{2 (\delta t/\Delta t) m_{\rm DM}^2}{K_{\rm X,S} \kappa^2 \rho_{\rm DM}} = 1 \times 10^{32} \l \frac{m_{\rm DM}}{10^{-6} \text{ eV}} \r^2 \l \frac{5}{K_{\rm X,S}} \r  \l \frac{\delta t/\Delta t}{7\times 10^{-17}} \r, 
\eea
where $K_{\rm X,S}$ is the enhancement factor of the space-based clock, $\rho_{\rm DM} = 0.4 \, \rm GeV/cm^3$ and $m_{\rm DM} \gg 10^{-9} \rm \, eV$, where $a_0(h,m_{\rm DM}) = 1$ for the ISS orbit. In the low mass regime, the value of $a_0$ is suppressed, $a_0 = \l h /R_\oplus\r^2$, as we have shown in Eq. \eqref{eq:phiSqApprox}. In this case the bound on the coupling becomes
\bea \label{eq: dx phase low mass}
d_{X}^{(2)} \lesssim \l\frac{R_\oplus}{h}\r^2\frac{2(\delta t/\Delta t) m_{\rm DM}^2}{K_{\rm X,S} \kappa^2 \rho_{\rm DM}} = 3 \times 10^{22} \l \frac{m_{\rm DM}}{10^{-12} \text{ eV}} \r^2 \l \frac{5}{K_{\rm X,S}} \r  \l \frac{\delta t/\Delta t}{7\times 10^{-17}} \r \l \frac{0.06}{h/ R_\oplus} \r^2.
\eea

 \subsubsection{Frequency measurement}
Looking for a change in phase is often superior to searching for a direct frequency shift.  This is because one can compare the phase of a space-based clock and an Earth-based one at the beginning and end of an integration period; this integrates over the modified frequency for the entirety of the experiment. In contrast, a direct frequency comparison requires establishing a link between clocks, so that the frequency can be compared at a given time.  This limits the integration time to the period when the space- and ground-based clocks are in visual contact. Depending on the type of orbit, the total time of visual contact can be 1-2 orders of magnitude shorter than the full duration of the experiment. 

So, whenever keeping track of the phase is possible for the duration of the experiment (as is the case for the Cesium fountain clock used in the ACES mission), it provides the best performance when looking for long-term linear drifts. However, when optical clocks are considered, achieving phase coherence for periods exceeding several hours is challenging \cite{Derevianko:2021kye}. In this case, it may be of interest to instead perform a frequency measurement.

As discussed in Sec.~\ref{sec: cos thetar}, when comparing frequencies between space- and Earth-based clocks, the location of the latter impacts the sampling of $\cos \theta_r$, see Eq. \eqref{eq: costheta 2}. This expression implies that space-to-ground comparisons do not sample the full range of values of  $\cos \theta_r$.  Moreover, these comparisons introduce a constant offset $\langle \cos \theta_r \rangle = \sin  \delta_{\rm DM}  \sin   \delta_{\rm lab} $. The observed frequency shift is given by
\bea
\Delta f_{\rm DM}(t_{\pm,n}) &=&  \Delta f_{\rm DM,0} \sum_{l = 0}  a_l (h,m_{\rm DM}) P_l \l \cos \theta_r (t_{\pm,n}) \r \nonumber \\
&=& \Delta f_{\rm DM,0} \sum_{l = 0}  \beta_l (h,m_{\rm DM}) P_l 
\l \cos \l \frac{\Omega}{\omega} 2\pi n  \pm \alpha_0\r \r,
\eea
where $\beta_0  = a_0 (h,m_{\rm DM}) + a_1 (h,m_{\rm DM}) \sin{\delta_{\rm DM}} \sin {\delta_{\rm lab}} + \mathcal{O}(a_2(h,m_{\rm DM})) $ represents the 
effective monopole, and $\beta_1 = a_1(h,m_{\rm DM}) \cos { \delta_{\rm DM} }  \cos {\delta_{\rm lab}}+ \mathcal{O}(a_2(h,m_{\rm DM})) $ parametrizes the leading temporal dependence, where
 the time dependence enters via the $n$.

Note, the frequency shift depends on the latitude where the Earth-based clock is located. Since both $a_1$ and $\delta_{\rm DM}$ are negative, to maximize the static signal ($\beta_0$) the clock should be located in the Northern Hemisphere, at latitudes where it can observe the clock reaching its peak declination, $\delta_{\rm lab} \approx \delta_0$. 
To maximize the amplitude of $\beta_1$, the Earth-based clocks should be located where $\cos \delta_{\rm lab} = 1$, i.e. at the equator. 
The measurement accuracy of the \emph{static} frequency shift, $\langle \Delta f_{\rm DM}(t)\rangle$,
is limited by the clock's
fractional frequency uncertainty 
\bea \label{eq: freq comp lin}
\frac{\langle \Delta f_{\rm DM}(t)\rangle}{f_0} \approx \beta_0(h,m_{\rm DM}) \frac{\Delta f_{\rm DM,0}}{f_0} \lesssim   \sqrt{\sigma_{y}^2(\tau) + \sigma_0^2}.
\eea
Here $\tau$ represents the measurement time, expressed as a dimensionless count of one-second intervals; $\sigma_y (\tau) \propto \tau^{-1/2}$ is the short-term stability of the clock. 
For long integration times the clock's fractional frequency uncertainty saturates at the noise floor $\sigma_0$. 

A typical Earth-based clock is located in the Northern Hemisphere around the optimal location for measuring the static signal, where $\sin  \delta_{\rm lab} \sin  \delta_{\rm DM}  \approx 0.5$.  This leads to a sensitivity in the $m_{\rm DM} \gg 10^{-9} \, \rm eV$ limit:
\bea \label{eq: dx freq const}
d_{X}^{(2)} \lesssim \frac{2\sigma_0 m_{\rm DM}^2}{\beta_0 K_{\rm X,S} \kappa^2 \rho_{\rm DM}} = 2 \times 10^{29} \l \frac{m_{\rm DM}}{10^{-6} \text{ eV}} \r^2 \l \frac{10}{K_{\rm X,S}} \r  \l \frac{\sigma_0}{2.4 \times 10^{-19}} \r \l \frac{1.2}{\beta_0} \r.
\eea
In the low mass limit $\beta_{0} \approx a_0 \approx \l h / R_\oplus\r^2$, and 
\bea \label{eq: dx freq const low}
d_{X}^{(2)} \lesssim \l\frac{R_\oplus}{h}\r^2 \frac{2\sigma_0 m_{\rm DM}^2}{K_{\rm X,S} \kappa^2 \rho_{\rm DM}} = 5 \times 10^{19} \l \frac{m_{\rm DM}}{10^{-12} \text{ eV}} \r^2 \l \frac{10}{K_{\rm X,S}} \r  \l \frac{\sigma_0}{2.4 \times 10^{-19}} \r \l \frac{0.06}{h/ R_\oplus} \r^2.
\eea
In both equations we have assumed an integration time long enough to reach the noise floor. If this condition is not met, one should substitute $\sigma_0 \to \sigma_y (T_{\rm int}/1 \, \rm s)$ in the above expressions.

To estimate the sensitivity of the experiment to frequency \emph{modulation}, we assume  that each measurement lasts a time $\tau$ and that the totality of the experiment runs over a long enough period $T_{\rm int} = N_{\rm meas} \tau$ to measure at least a half-cycle of orbital precession. The fractional frequency uncertainty of an individual measurement is dominated by the short-term stability $\sigma_y(\tau)$. Then, the signal-to-noise ratio (SNR) is given by

\bea \label{eq: SNR s-g}
\text{SNR}^2 &=& \sum_{k = 1}^{N_{\rm meas}} \frac{(\frac{\Delta f_{\rm DM} (t_k)}{f_0} -  \langle \frac{\Delta f_{\rm DM}(t) }{f_0} \rangle)^2}{\sigma_{y}^{2}(\tau)} \\ 
&=& \frac{N_{\rm meas}}{\sigma_{y}^{2}(\tau)} \l \frac{\Delta  f_{\rm DM,0}}{f_0}\r^2 
\l \sum_{l, l'} \beta_l \beta_{l'} \langle P_l (\cos \eta) P_{l'} (\cos \eta) \rangle_\eta -  \l \sum_{l} \beta_l  \langle P_l (\cos \eta) \rangle_\eta \r^2 \r  \nonumber\\
&\approx& \frac{1}{\sigma_{y}^{2}(T_{\rm int}/1\rm \, s)} \l \frac{\Delta  f_{\rm DM,0}}{f_0}\r^2 \frac{\beta_1^2}{2} + \mathcal{O}(\beta_2^2, \beta_1 \beta_3),\nonumber
\eea
where $\langle ... \rangle_\eta = (2\pi)^{-1} \int^{2\pi}_{0} d \eta$ represents an average over phases. The second line takes  advantage of the fact that the ISS orbit finely samples the oscillatory signal: the signal period is roughly 72 days, and even a single Earth-based clock allows measurements to be performed 5-6 times a day. The third line uses the scaling of the short-term stability of the clock.  Setting $\rm SNR = 1$, and making similar assumptions about the clock's location as in Eq. \eqref{eq: dx freq const}, we can again write down sensitivities in the low and high-mass regimes.  For $m_{\rm DM} \gg 10^{-9} \, \rm eV$, we have

\bea \label{eq: dx freq osc}
d_{X}^{(2)} \lesssim \frac{2\sqrt{2}\sigma_y(T_{\rm int}/1 \, \rm s) m_{\rm DM}^2}{\vert \beta_1 \vert K_{\rm X,S} \kappa^2 \rho_{\rm DM}} = 2 \times 10^{29} \l \frac{m_{\rm DM}}{10^{-6} \text{ eV}} \r^2 \l \frac{10}{K_{\rm X,S}} \r \l \frac{0.2}{\vert \beta_1 \vert} \r \l \frac{\sigma_y(1)}{ 10^{-16}} \r \l \frac{ 10^6 \, \rm s}{T_{\rm int}} \r^{\frac{1}{2}}.
\eea
In the low mass limit, $\beta_1 \approx  C (k_0 h)^2 \cos \delta_{\rm DM}   \cos  \delta_{\rm lab}  $, and we have
\bea \label{eq: dx freq osc low mass}
d_{X}^{(2)} &\lesssim& \frac{2\sqrt{2}\sigma_y(T_{\rm int}/ 1 \, \rm s) }{C  v_{\rm DM}^2 h^2 \cos \delta_{\rm DM}   \cos  \delta_{\rm lab}  K_{\rm X,S} \kappa^2 \rho_{\rm DM}} \\
&=& 9 \times 10^{21}  \l \frac{10}{K_{\rm X,S}} \r \l \frac{0.06 R_{\rm \oplus}}{h} \r^2 \l \frac{0.5}{\cos  \delta_{\rm DM}   \cos  \delta_{\rm lab}  } \r \l \frac{\sigma_y(1)}{ 10^{-16}} \r \l \frac{ 10^6 \, \rm s}{T_{\rm int}} \r^{\frac{1}{2}},
\eea
where we assumed $v_{\rm DM} = 233 \, \rm km/s$ \cite{Catena:2011kv}.

Since a static frequency shift and its modulation are related, the overall limit is set by the more competitive of the two. The search for the former provides a more powerful constraint on $d_{X}^{(2)}$ than a measurement of the latter by a factor of $\beta_0/\beta_1$ as long as the total integration time is shorter than the time required to reach the noise floor, $\sigma_y(T_{\rm sat}/1 \, \rm s) \simeq \sigma_0 $.
 Eventually, if the total integration time is greater than $T_{\rm int} \gtrsim (\beta_0/\beta_1)^2 \, T_{\rm sat}$, the time-dependent signal leads to stronger constraints. The dipole-induced (i.e. modulated) signal also offers a way to test whether a hypothetical static frequency shift originates from quadratically coupled scalar DM. In particular, the phase of dipole-induced oscillation depends on the direction of the DM wind, the orientation of the space clock's orbit, and the location of the Earth-based clock, so the dipole signal must not only be present but also have the correct phase. Furthermore, if a monopole-like shift and a dipole component are both detected, their ratio $a_1/a_0$ could, in principle, be used to identify the approximate DM mass.

\subsection{Space-to-space experiments}
\label{sec:Space to Space}
Another interesting possibility is to co-locate a pair of clocks in space \cite{Tsai:2021lly}. In this setup both clocks experience the same DM background. However, if their sensitivity to the change in fundamental parameters is different $K_{X,1}\neq K_{X,2}$, see Eq.~(\ref{eq: parameter to freq}), their responses to changing DM background will differ, leading to a change in their frequency ratio. In particular

\bea
\frac{\Delta (f_1/f_2)}{(f_1/f_2)} \approx \frac{\Delta f_1}{f_1} - \frac{\Delta f_2}{f_2} = 
\frac{\Delta K_{X}  d^{(2)}_X \kappa^2 \phi^2(\Vec{x})}{2}, 
\eea
where we defined $\Delta K_{X} = K_{X,1} - K_{X,2}$.
To maximize a dipole signal, it is beneficial to operate the experiment in the region where $a_1(h,m_{\rm DM}) \sim {\mathcal O}(1)$. Similar to the previous section, we assume the experiment operates on the ISS orbit, where this condition is satisfied for $m_{\rm DM} \gtrsim 10^{-9} \, \rm eV$. 

A benefit of performing a measurement with co-located clocks is the ability to continuously compare them. This significantly reduces dead time compared to the space-to-ground frequency comparison.

The signal inherits its   temporal dependence from the time dependence of $\cos \theta_r(t)$ as
\bea
\chi(t) \equiv \frac{\Delta (f_1/f_2)}{(f_1/f_2)} &=& \Delta K_{X} \frac{d^{(2)}_X \kappa^2 \phi_{0}^2}{4} \sum_{l=0}^{N} a_l (h,m_{\rm DM}) P_l(\cos \theta_r (t)) \\
&=& \frac{\Delta \Tilde{f}}{\Tilde{f}} \sum_{l=0}^{N} a_l (h,m_{\rm DM}) P_l(\cos \theta_r (t)).
\eea
where we defined $\Delta \Tilde{f}/\Tilde{f} = \Delta K_{X} d^{(2)}_X \kappa^2 \phi_{0}^2/4 $. The time dependence of $\cos \theta_r(t)$ is given in Eq.~(\ref{eq: costheta}). In contrast to the space-to-ground comparison, the frequency of this signal coincides with the frequency of the orbit, which could potentially complicate extracting the signal from the noise -- there could well be  sources of noise that share this frequency. However, the phase and amplitude of the signal are determined by the satellite's position and the direction of the DM wind. This should help to distinguish a DM-induced signal from other effects.

To estimate the sensitivity to this signal,  we again calculate a SNR assuming a series of $N_{\rm meas}$ measurements, each lasting $\tau$ seconds, amounting to a total integration time of $T_{\rm int}/1 \; \rm{s} = N_{\rm meas} \tau$, with the accuracy of a single measurement given by the short-term stability $\sigma_y(\tau)$.  We find: 
\bea\label{eq:SNR}
\text{SNR}^2 &=& \sum_{k = 1}^{N_{\rm meas}} \frac{(\chi (t_k) -  \langle \chi(t) \rangle)^2}{\sigma_{y}^{2}(\tau)} \\ 
&=& \frac{N_{\rm meas}}{\sigma_{y}^{2}(\tau)} \l \frac{\Delta \Tilde{f}}{\Tilde{f}}\r^2 
\l \sum_{l, l'} a_l a_{l'} \langle P_l (\cos \theta_r(t)) P_{l'} (\cos \theta_r(t)) \rangle_t -  \l \sum_{l} a_l  \langle P_l (\cos \theta_r(t)) \rangle_t \r^2 \r  \nonumber\\
&\approx& \frac{1}{\sigma_{y}^{2}(T_{\rm int}/1 \, \rm s)} \l \frac{\Delta \Tilde{f}}{\Tilde{f}}\r^2 \frac{a_1^2}{3} + \mathcal{O}(a_2^2, a_1 a_3),\nonumber
\eea
%%%
where $\langle ... \rangle_t = T_{\rm int}^{-1} \int^{T_{\rm int}}_{0} d t$ represents averaging over the duration of the experiment. When performing this average we assumed orbital parameters of the ISS, $\delta_0 = 52 ^{\circ}$ and $\omega/\Omega \approx 1200$. For dramatically different orbits, for example, those that are nearly circumpolar, the numerical coefficient of the $a_1^2$ term above would increase by roughly 30\%.

To translate Eq.~\eqref{eq:SNR} into a sensitivity projection, we require $\text{SNR} = 1$, which yields 
\bea \label{eq: dx co-located}
d_{X}^{(2)} \lesssim  \frac{2 \sqrt{3} \sigma_y(T_{\rm int}/1 \, \rm s) m_{\rm DM}^2}{\rho_{\rm DM} \vert a_1 \vert \Delta K_X \kappa^2} = 3 \times 10^{29} \l \frac{m}{10^{-6} \text{ eV}} \r^2 \l \frac{10^6 \, \rm s }{T_{\rm int}} \r^{\frac{1}{2}} \l \frac{\sigma_y(1)}{10^{-16}} \r \l \frac{0.4}{\vert a_1 \vert} \r \l \frac{10}{\Delta K} \r.
\eea 
If we specialize to the low mass limit where $a_1 =  C (k_0 h)^2 $, we find:
\bea \label{eq: dx co-located low mass}
d_{X}^{(2)} \lesssim  \frac{2 \sqrt{3} \sigma_y(T_{\rm int}/ 1 \, \rm s) }{C \rho_{\rm DM} v_{\rm DM}^2 h^2 \Delta K_X \kappa^2} = 6 \times 10^{21}  \l \frac{10^6 \, \rm s }{T_{\rm int}} \r^{\frac{1}{2}} \l \frac{\sigma_y(1)}{10^{-16}} \r \l \frac{0.06}{h/ R_\oplus} \r^2 \l \frac{10}{\Delta K} \r.
\eea 

A measurement of a time-dependent frequency shift with co-located clocks leads to parametrically the same constraints (up to $\mathcal{O}(1)$ factors) as the equivalent measurement with space-to-ground comparisons, see Eqs. \eqref{eq: dx freq osc} and \eqref{eq: dx freq osc low mass}. However, co-located clocks can achieve much longer integration times, so ultimately may be more sensitive.

\subsection{Sensitivity projections}

\begin{table}[t] 
\centering
 \begin{tabular}{||c | c c c c c c||} 
 \hline
  Experiment & $\sigma_y(\tau)$ & $\sigma_0$ & $\vert K_\alpha \vert$ & $\vert K_\mu \vert$ & $\vert K_q \vert$ & $\vert K_N \vert$ \\ [0.5ex] 
 \hline
  ACES$^*$ & $1 \times 10^{-13}/\sqrt{\tau}$ & $1 \times 10^{-16}$ & $5$ & $2$ & $0$ & $1$\\ 
 optical & $1\times 10^{-16}/\sqrt{\tau} $  & $2.4 \times 10^{-19}$ & $10$ & $1$ & $0$ & $0$ \\
 nuclear  & $1\times 10^{-15}/\sqrt{\tau} $ &  $1 \times 10^{-19}$ & $10^4$ & $0$ & $10^5$ & $10^5$\\
  optical - optical & $1\times 10^{-16}/\sqrt{\tau} $  & $2.4 \times 10^{-19}$ & $10$ & $0$ & $0$ & $0$ \\
 nuclear - optical  & $1\times 10^{-15}/\sqrt{\tau} $ &  $1 \times 10^{-19}$ & $10^4$ & $1$ & $10^5$ & $10^5$\\
 [1ex] 
 \hline
 \end{tabular}
 \caption{Parameters used to estimate experimental sensitivity. The short-term ($\sigma_y(\tau)$) and long-term ($\sigma_0$) stability of each clock type are given in columns 2 and 3. In the former, $\tau$ represents the measurement time,  expressed as a dimensionless count of one-second intervals.  The $ K_X $ with $X \in (\alpha,\mu,q , N )$ represent enhancement factors for the clocks' sensitivity to variations of the fine-structure constant, electron mass, average light-quark mass, and the nucleon mass respectively, see Eq.~(\ref{eq: parameter to freq}). The ACES mission \cite{Savalle:2019isy} is marked with an asterisk as it measures the phase drift rather than a frequency ratio. The mission plans to reach a long-term frequency stability at the $10^{-16}$ level, which translates to measuring the phase drift at the $7 \times 10^{-17}$ level. }
 \label{tab: experimental parameters}
\end{table}

In this section we apply the above sensitivity estimates to various experimental setups involving quantum clocks. Benchmark experimental parameters are given in Table \ref{tab: experimental parameters}. We start with a summary of types of clocks considered, their sensitivity to fundamental parameters, and frequency resolution. We consider three distinct types of clocks: a Cesium fountain clock, an optical clock, and a single-ion Thorium clock.

A Cesium fountain clock is based on a hyperfine transition at microwave frequencies.  As the fractional frequency uncertainty of the clock is inversely proportional to its frequency, a Cesium clock has significantly lower sensitivity than optical or nuclear clocks. Because the transition depends on interaction between the electron and nucleus,  the Cesium transition is sensitive to changes in the electron mass, fine structure constant, and nucleus mass. In our sensitivity estimates we take clock parameters from the ACES mission \cite{Savalle:2019isy}.

A typical optical transition is sensitive to electron mass and fine-structure constant. In this case, the transition energy is proportional to Rydberg's constant, $\text{Ry} \propto m_e \alpha^2 $.  Thus, one might expect $\Delta f/f \sim 2 \, \delta \alpha/\alpha$, i.e. $K_{\alpha} \sim 2$.  However, there are transitions where the sensitivity to the fine-structure constant is enhanced by relativistic effects to $K_{\alpha} \sim \mathcal{O}(10)$ \cite{Safronova:2018quw}, which is what we assume in our sensitivity projections. To estimate the sensitivity of a space-borne optical clock system we take mission requirements from FOCOS proposal \cite{Derevianko:2021kye}.  

The final type of clock we consider is a single-ion Thorium clock. Thorium has a nuclear transition with energy $8 \, \rm eV$ \cite{ Zhang:2024ngu, Tiedau:2024obk}.  This makes it special -- typical nuclear transition energies are many orders of magnitude greater. The small energy associated with this transition results from a relatively fine cancellation between the electromagnetic and nuclear contributions to the transition energy.   This leads to an enhanced sensitivity to the fine structure constant $K_\alpha\sim 10^4$, nucleon masses $K_N \sim 10^5$ and quark masses $K_q \sim 10^5$ \cite{Beeks:2024xnc,Caputo:2024doz}. As a performance estimate, we take a well-known estimate for a single ion clock \cite{Campbell:2012zzb}.

When combining clocks into two-clock systems we consider two options: i) two optical clocks and ii) one nuclear clock with an optical clock as a reference. In the first configuration (optical-optical), we assume that two optical transitions are chosen such that their relative sensitivity to the fine-structure constant is enhanced by $K_{\alpha,1} - K_{\alpha,2} \sim \mathcal{O}(10)$. The sensitivity to the electron mass for any optical transition is of the form $K_{\mu} = 1+ \epsilon$, where $\epsilon\ll1$. Thus, the relative sensitivity of the optical-optical setup to the electron mass is $K_{\mu,1} - K_{\mu,2} \sim \mathcal{O}(\epsilon)$  and we do not discuss it further here. In the second configuration (nuclear - optical), the clock system inherits the sensitivity of the nuclear clock to the fine-structure constant, gluons, and average quark mass while adding sensitivity to electrons through the optical clock. The overall frequency resolution of this system is limited by the performance of the nuclear clock.

To determine the clock's response requires the dark matter profile, which in principle depends on both the DM mass and coupling.  
However, for the couplings of interest, the field profile is well-approximated by imposing a Dirichlet boundary condition, i.e. it does not depend on the exact value of the coupling. Consequently, we need only perform a scan over the DM mass. For each mass we compute the corresponding $a_l$ coefficients up to $l\leq 3$. We use these coefficients to determine the coupling strength that saturates sensitivity for either i) static signals in phase measurements (Eq. \eqref{eq: time res}), ii) static signal in frequency measurements (Eq. \eqref{eq: freq comp lin}) or iii) a signal-to-noise ratio of $\text{SNR} = 1$ when searching for time-varying signals in space-to-ground (Eq. \eqref{eq: SNR s-g}) or space-to-space comparisons (Eq. \eqref{eq:SNR}). Where applicable we assume that the Earth-based clock is located in the Northern Hemisphere, such that $\sin  \delta_{\rm lab} \sin  \delta_{\rm DM}  = 0.5$. For masses above $m_{\rm DM} > 4 \times 10^{-7} \, \rm eV$, we use the asymptotic values of $a_l$ coefficients listed in Table \ref{tab: legendre coefficients}. For all clock experiments (except for FOCOS) we assume $h = 400 \, \rm km$, corresponding to the approximate orbit height of the ISS.\footnote{In practice, the effective distance to the scattering surface depends on the coupling strength, because stronger couplings allow higher atmospheric layers to reflect the scalar field. The effect is largest for the least sensitive experiments. For ACES it can shift $h$ by up to $150 \, \rm km$ in the high mass limit. While this seems like a significant effect, it mostly affects the anisotropic component of the signal (to which ACES is not sensitive). Even in that case, its impact on the dipole term is small, as $a_1$ at $h = 250 \, \rm km$ and $h = 400 \, \rm km$ only differs at the $\mathcal{O}(10 \%)$ level.  The effect would mildly increase the signal strength. Another consequence is that the transition to low mass limit starts at a slightly higher mass. These effects are sub-leading, so we ignore them for simplicity.    }

When projecting sensitivities we use the clock parameters in Table ~\ref{tab: experimental parameters}. For an experimental configuration that searches for a phase drift (ACES), we assume a relative phase uncertainty of $\delta t/\Delta t = 7 \times10^{-17}$ \cite{Savalle:2019isy}.  For other space-to-ground missions, the integration time is limited by the time spent in visual contact. In a one-clock-in-space and  one-clock-on-Earth setup, this reduces the effective integration time by about two orders of magnitude. We set $T_{\rm int} = 10^6 \, \rm s$, i.e., a mission duration of $10^8 \, \rm s$. Therefore, for searches for a constant frequency offset, we approximate the total fractional frequency uncertainty $\sigma\simeq\max (\sigma_0, \sigma_y (10^6))$. 
When looking for an oscillatory signal with space-to-ground comparisons, see Eq.~(\ref{eq: SNR s-g}), we assume matched filtering is applied.  In principle, this allows experiments to reach a fractional frequency uncertainty below $\sigma_0$.  To show how this could occur, we set $\sigma \simeq \sigma_y (\tau)$ combined with an extended integration time $T_{\rm int} = 10^{7} \, \rm s$.  Such a long integration time  would likely require ${\mathcal{O}(10)}$ clocks on Earth to decrease the dead time. 

\begin{figure}
    \centering
    \includegraphics[width=\columnwidth]{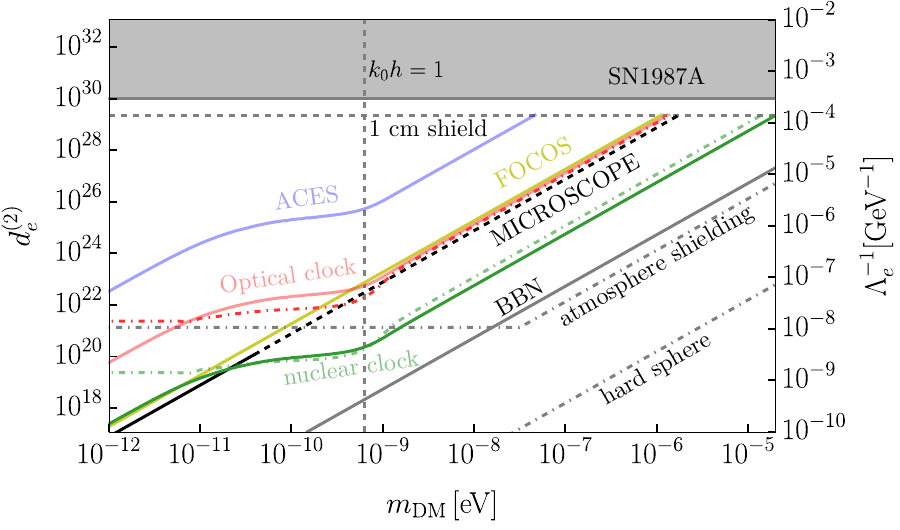}
    \caption{Sensitivity projections for space-to-ground comparisons for a quadratic coupling to photons as a function of dark matter mass. Solid colored lines represent searches for a linear drift in phase or a constant offset in frequency, while dot-dashed lines represent searches for an oscillatory signal. The colored lines depict sensitivities of the ACES mission (blue) \cite{Savalle:2019isy}, an optical clock on the ISS (red), a nuclear clock on the ISS (green), and the FOCOS mission (yellow) \cite{Derevianko:2021kye}, respectively.  Assumed experimental parameters are listed in Table \ref{tab: experimental parameters}. For the fractional phase uncertainty of the ACES mission, we assume $\delta t/\Delta t = 7 \times 10^{-17}$. For searches for a constant frequency offset, we take the total fractional frequency uncertainty to be $\max (\sigma_0, \sigma_y (10^6))$, while for oscillatory-signal searches we use the short-term stability $ \sigma_y (10^7)$. The black line gives the bound from MICROSCOPE \cite{MICROSCOPE:2022doy,Banerjee:2022sqg}, and the dashing indicates the range of masses where the low momentum approximation used in Ref.~\cite{Banerjee:2022sqg} breaks down \cite{Gan:2025nlu}, see text for details. The solid gray lines represent bounds from SN \cite{Gan:2025nlu} and BBN \cite{Bouley:2022eer}, while dashed gray lines indicate boundaries where reflection from the atmosphere or satellite shielding becomes important.  The vertical dashed gray line ($k_0 h =1 $) depicts the transition from low mass regime to high mass regime for $400 \, \rm km$ orbit.  
}
    \label{fig: space ground photon}
\end{figure}
In Fig. \ref{fig: space ground photon}, we depict the sensitivity of space-to-ground comparisons to the photon coupling $d^{(2)}_e$. In this case, the clock (network) on the ground serves as a reference (the DM signal is screened), while the space-based clock experiences a non-trivial DM profile. 
Here, solid lines represent projected limits arising from measurements of constant phase drift or constant frequency displacement. 
Dot-dashed lines represent searches for time-varying signals, due to the field profile not being spherically symmetric. Blue, red, and green lines represent constraints from microwave (i.e. ACES), optical, and nuclear clocks on the ISS, respectively. The yellow line represents the scenario described in the FOCOS proposal \cite{Derevianko:2021kye}, where the clock is sufficiently far from Earth that the local DM density is unaffected by Earth-DM scattering. Here, we have assumed that the satellite carries an optical clock with parameters from Table \ref{tab: experimental parameters}, consistent with the original proposal of Ref.~\cite{Derevianko:2021kye}.  For some values of $m_{DM}$, especially near $10^{-9}$ eV, the dashed-red line extends to lower values of $d_{e}^{(2)}$ than the yellow line.  Since the two lines are derived assuming identical clocks, this indicates the additional sensitivity that the dipole feature in the field provides.

\begin{figure}
    \centering
    \includegraphics[width=\columnwidth]{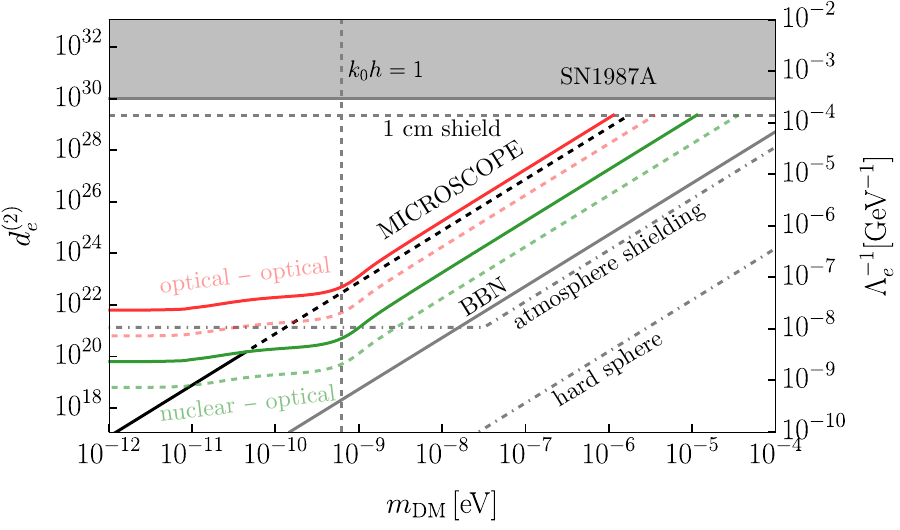}
    \caption{Sensitivity projections for space-to-space comparisons for a quadratic coupling to photons. Solid colored lines correspond to integration times of $10^6 \text{ s}$, while dashed lines represent integration times of $10^8 \text{ s}$. The colored lines depict sensitivities of a two optical clock system (red) and a nuclear - optical clock system (green). Remaining lines are as in Fig. \ref{fig: space ground photon}.  }
    \label{fig: space space photon}
\end{figure}

\begin{figure}
    \centering
    \includegraphics[width=\columnwidth]{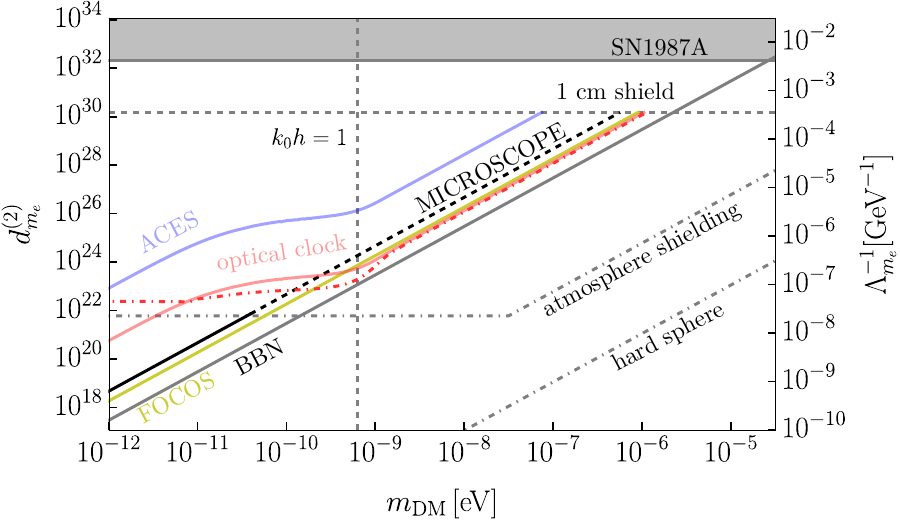}
    \caption{Sensitivity projections for space-to-ground comparisons for a quadratic coupling to electrons. Solid colored lines represent searches for a linear drift in phase or a constant offset in frequency, while dot-dashed lines represent searches for an oscillatory signal. For more details, see the caption of Fig. \ref{fig: space ground photon}.  }
    \label{fig: space space electron}
\end{figure}

\begin{figure}
\centering
    \includegraphics[width=\columnwidth]{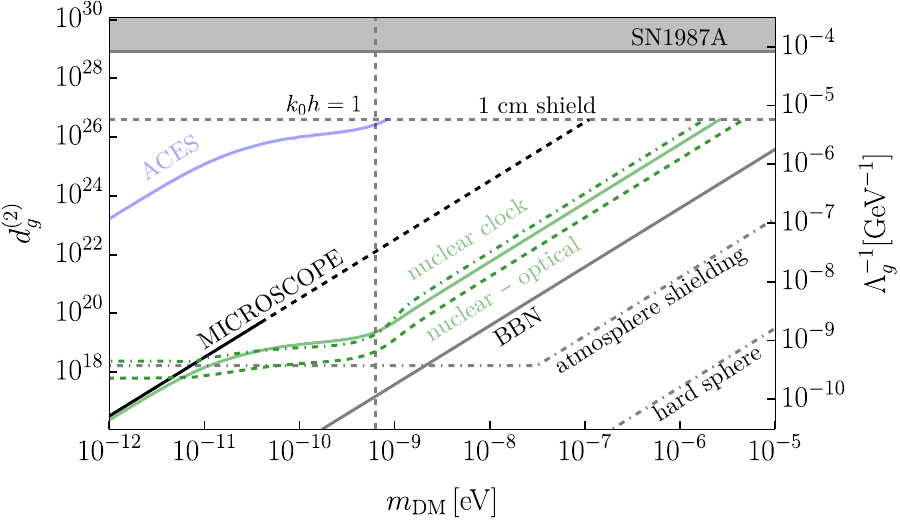}
    \label{fig: space space nucleon}
    \includegraphics[width=\columnwidth]{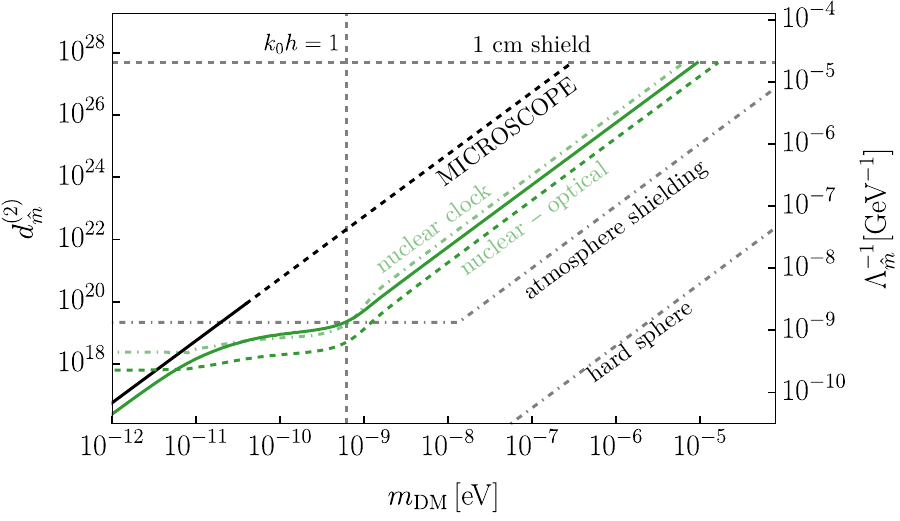}
    \caption{Sensitivity projections for space-to-ground and space-to-space comparisons for a quadratic coupling to gluons (top) and to quarks (bottom). The colored lines depict sensitivities for the ACES mission (blue) \cite{Savalle:2019isy}, a nuclear clock on the ISS referenced to a ground-based clock looking for a static (solid green) and oscillatory signal (dot-dashed green), and a nuclear--optical clock system co-located on the ISS (dashed green), respectively. Solid, dot-dashed and dashed green lines assume integration times of $10^6 \, \rm s$, $10^7 \, \rm s$ and $10^8 \, \rm s$, respectively. The remaining lines are the same as in Fig. \ref{fig: space ground photon}.}
    \label{fig: space space quarknucleon}
\end{figure}

In Figs. \ref{fig: space ground photon}, \ref{fig: space space photon}, \ref{fig: space space electron}, and \ref{fig: space space quarknucleon}, the gray region represents the SN1987A bound \cite{Olive:2007aj,Gan:2025nlu,Day:2023mkb}. In all figures, the black line represents the MICROSCOPE bound~\cite{MICROSCOPE:2022doy,Banerjee:2022sqg}. The latter bound is derived assuming a spherically symmetric force, following earlier work~\cite{Hees:2018fpg,Banerjee:2022sqg}, while adopting a $2\sigma$ uncertainty as in Ref.~\cite{Gan:2025nlu}. This assumption is only satisfied for 
lower mass values, $m_{DM} \lesssim 10^{-10}$ eV \cite{Gan:2025nlu}. Moreover, once $k_0 h \gtrsim 1$, the force is no longer dominated by the radial term and instead has direction approximately fixed towards the Cygnus constellation. We expect that a dedicated analysis that accounts for this anisotropy would change the bound by an $\mathcal{O}(1)$ factor compared to the one derived using the spherically symmetric ansatz. To indicate this uncertainty we use a dashed line in the regime where this anisotropy is significant. We find space-to-ground missions involving optical clocks should reach a comparable sensitivity on the $d_e$ coupling as the  MICROSCOPE mission.  Nuclear clocks would set leading constraints.  

In the geometric-optics limit, the multipole coefficients of the field profile reach their asymptotic values (Table \ref{tab: legendre coefficients}); this results in the observed scaling of the limits $\propto m_{\rm DM}^2$, see Eqs. \eqref{eq: dx phase}, \eqref{eq: dx freq const}, \eqref{eq: dx freq osc}, and \eqref{eq: dx co-located}. In the low-mass limit, searches for a phase drift or constant frequency offset are suppressed by the $\l h/R_\oplus\r^2$ factor \cite{Hees:2018fpg}, see Eqs. \eqref{eq: dx phase low mass} and \eqref{eq: dx freq const low}, while searches for an oscillating signal saturate as the combination $\phi_0^2 a_1$ reaches an asymptotic value, see Eqs. \eqref{eq: dx freq osc low mass} and \eqref{eq: dx co-located low mass}.

At very large couplings, the satellite itself could act to attenuate the dark matter.  We assume there is approximately $1$ cm of aluminum material outside the clock \cite{ESABASE2Shielding}.  The line where the attenuation length $\lambda \approx \sqrt{d_X^{(2)} \kappa^2 Q_{\rm shield}\rho_{\rm shield}}$  is less than 1 cm is shown as a dashed horizontal line in Figs.~\ref{fig: space ground photon}, \ref{fig: space space photon}, \ref{fig: space space electron}, and \ref{fig: space space quarknucleon}. 
Additionally, vertical gray dashed lines labeled $k_0 h = 1$ depict the transition from the low-mass to high-mass regime, while the gray solid line depicts the bound from BBN \cite{Bouley:2022eer} that assumes an extrapolation of the field value back to that era. Finally, gray dot-dashed lines represent approximate boundaries above which the scalar field profile can be well-approximated by imposing a Dirichlet boundary condition at the Earth's surface (hard sphere) or at the altitude at which atmosphere reflects the field (atmospheric shielding). The shape of these boundaries is determined by requiring $\Delta m_{\rm DM, atm}^2 \gtrsim k^2_0$ ($\Delta m_{\rm DM, \oplus}^2 \gtrsim k^2_0$) and $\sqrt{\Delta m_{\rm DM, atm}^2} h_{\rm atm} \gtrsim 1$ ($\sqrt{\Delta m_{\rm DM, \oplus}^2} R_{\oplus} \gtrsim 1$), where $h_{\rm atm}\approx 8 \, \rm km$ is the effective thickness of the dense part of the atmosphere. The parameters used to estimate these boundaries are given in Table \ref{tab: surface parameters}.

\begin{table}[t] 
\centering
 \begin{tabular}{||c | c c c c c||} 
 \hline
  location & $\rho \, [\rm g/cm^3]$  & $Q_e$ & $Q_{m_e}$ & $Q_{g}$ & $Q_{\hat{m}}$  \\ [0.5ex] 
 \hline
  lower atmosphere & $ 10^{-3}$ & $1.3 \times 10^{-3}$ & $2.7 \times 10^{-4}$ & $1$ & $7.8 \times 10^{-2}$ \\ 
 Earth's surface & $5.5$  & $2.0 \times 10^{-3}$ & $2.7 \times 10^{-4}$ & $1$ & $8.1 \times 10^{-2}$  \\
 %[1ex] 
 \hline
 \end{tabular}
 \caption{Parameters used to estimate boundaries of atmospheric shielding and hard scattering from the Earth's surface. Density $\rho$ of a given medium is given in the second column. Other columns indicate the dilatonic charges $Q_X$, see Eq. \eqref{eq: dilatonic charge}, to photons, electron mass, gluons and average light quark mass respectively.}
 \label{tab: surface parameters}
\end{table}

In Fig. \ref{fig: space space photon}, we show the sensitivity curves corresponding to two co-located clocks in space. 
Here, we also assume an ISS-like orbit.  In this case, the signal arises from the non-spherically symmetric field profile. Solid lines assume an integration time of $10^6\, \rm s$, while dashed lines assume an integration time of $10^8 \, \rm s$. The red lines assume a two-clock system using two different optical transitions; the green line assumes a nuclear-optical clock system where the optical clock is used as a frequency reference (as $K_{\alpha,\, \rm nuclear} \gg K_{\alpha,\, \rm optical}$). For integration times of $10^6 \, \rm s$ in the high-mass regime, we recover a similar sensitivity as space-to-ground missions. However, if the clocks run for $10^8 \, \rm s$, the double optical clock system could slightly improve upon the sensitivity of MICROSCOPE, and a nuclear-optical pair should probe significant parameter space inaccessible to lab-based experiments.

In the remaining figures we display   sensitivity curves for the electron coupling $d^{(2)}_{m_e}$ (Fig. \ref{fig: space space electron}) and the gluon and quark couplings $d^{(2)}_{g}$ and $d^{(2)}_{\hat{m}}$ (Fig. \ref{fig: space space quarknucleon}). The color schemes for these figures are as in  Figs.~\ref{fig: space ground photon} and~\ref{fig: space space photon}.
For $d^{(2)}_{m_e}$ (Fig. \ref{fig: space space electron}) optical clocks can slightly improve upon the MICROSCOPE sensitivity and get within a factor of 2 of the BBN bound, but since no clocks have an enhanced sensitivity to $d_{m_e}$ coupling, reaching the atmospheric ceiling seems unlikely. While a nuclear-optical system could in principle be used to probe this coupling, even with $T_{\rm int} = 10^8 \, \rm s$ the  sensitivity is comparable to an optical clock used in a space-to-ground comparison. So, for clarity, we omit the nuclear lines from the figure. On the other hand, searches for gluon and quark couplings (Fig. \ref{fig: space space quarknucleon}) can substantially benefit from nuclear clocks. In both cases a space-based nuclear clock referenced to a ground-based clock can significantly improve upon the MICROSCOPE mission by looking for either static or time-dependent signals (solid and dot-dashed green curves, respectively).  A co-located space-based nuclear-optical clock system offers an even better sensitivity (dashed green).

\clearpage

\newpage

\section{Conclusions}
\label{sec:conclusions}

We have investigated how quantum clocks in Low Earth Orbit can probe quadratically coupled ultralight scalar dark matter. 
In the regime where the field value is suppressed at the Earth's surface, its average galactic value is restored at distances sufficiently far from Earth. When the DM de Broglie wavelength is shorter than the Earth's radius, restoration of the field already happens at typical LEO altitudes, making them an ideal environment for probing the field. In the corresponding mass range, oscillations that are governed by the mass of DM are too rapid to be resolved by the clocks, but for quadratic couplings, the signal does not average out. Consequently, quantum clocks can be a valuable probe of the local DM density. 

Since the DM velocity distribution has a preferred direction, the field profile is not spherically symmetric and is instead well-approximated by a monopole plus a dipole term. As alluded to above, in LEO the monopole term can already be as strong as in deep space, which allows for efficient space-to-ground comparisons. However, the real benefit of LEOs is that the dipole amplitude is comparable to the monopole when orbits are at altitudes of a few hundred kilometers. The dipole feature can be translated to a time-dependent signal with orbital frequency and a well-known phase.  It can be probed with space-to-ground and co-located clock systems (or an orbital clock network \cite{Masia-Roig:2022net}). Measuring both the monopole and the dipole could provide a strong test of quadratically coupled scalar DM.  Moreover, if signals are observed, the ratio of these components could be used to estimate the DM mass.

Experiments performed at LEO are a natural testing ground for future High Earth Orbit \cite{Derevianko:2021kye} or deep-space missions \cite{Tsai:2021lly}.
The ongoing ACES mission at the ISS represents a first step towards probing quadratically-coupled DM in LEO with quantum clocks. While we found that ACES cannot improve on existing bounds from fifth-force searches, future iterations involving optical clocks may surpass the MICROSCOPE constraints for some DM-SM couplings. In addition, exploiting the dipole feature with a co-located nuclear-optical clock system could allow such missions to accumulate sufficient integration time to significantly strengthen the limits. 

The focus in this work has been on couplings that increase the DM mass in dense media. A similar analysis could be performed for couplings of opposite sign, although its validity is limited to regions where $m_{\rm DM,\oplus}^2>0$, which can be satisfied only for certain couplings and clock types. For some mass-coupling combinations the field should be non-negligible on Earth. To accurately characterize this case requires careful modeling of Earth's density profile, which is beyond the scope of this work. 

The approximate sensitivity of the MICROSCOPE mission used in this work (reproduced from Refs.~\cite{Hees:2018fpg,Banerjee:2022sqg}) assumes a spherically symmetric field profile. As already noted in \cite{Gan:2025nlu}, this assumption breaks down when the de Broglie wavelength of the DM becomes smaller than the size of the Earth. That work demonstrated the resulting anisotropy by comparing the force in the forward and backward directions. A complete analysis to compute a robust MICROSCOPE bound would require a detailed characterization of the anisotropy and understanding how it translates to observables.
Our work provides the angular dependence of the field profile, an important step towards such a full analysis, which will be presented elsewhere.

\section*{Acknowledgments}
D.B. and A.P. are supported by
the Department of Energy under grant number DE-SC0007859. D.B. is supported in part by a Leinweber postdoctoral fellowship.

\appendix
\section{Multipole expansion} \label{App: multipole expansion}
In this Appendix, we present results that are useful in the  derivation of the scalar field profile. The goal is  technology that provides efficient access to the $a_{l}$ coefficients of Eq.~(\ref{eq: phi^2 legendre}). 
These coefficients are of immediate use in the clock analysis of this paper, but since the field profile acts as a potential for the background-induced force (wake force), the multipole expansion also provides an efficient framework to describe such forces in a vicinity of spherically symmetric bodies.  

To find the field value numerically at a given location using the methods of Sec.~\ref{sec:profile}, one must both integrate over the momentum distribution and sum over the contributing multipoles.  This procedure can be numerically expensive. In this Appendix we analytically perform the integrations over angular variables.  The problem reduces to a one-dimensional integral and a sum over multipoles, which reduces the complexity of the numerical computation. 

The multipole coefficients capture the angular dependence of the field. We assume only that the field can be decomposed into states of definite angular momentum. The results can thus be applied to settings beyond those considered in this work, such as perturbative scattering, the case of an opposite-sign coupling, or scattering from a sphere with a radially varying density $\rho(r)$. 

Decomposing the field $\psi$ into states of definite angular momentum gives 
\bea
\psi = \vert \psi_0 \vert \sum_{l}(2l+1)i^l  P_l \l \cos \theta \r F_l(k, d_X^{(2)}, r)
\eea
The $F_{l}$ encode the solution to the quantum mechanical scattering problem, see Eq.~(\ref{eq:multipole}).
Squaring the field yields
\bea
\label{eq:psisqApp}
\vert \psi \vert^2 = \vert \psi_0 \vert^2 \sum_{l,l'}(2l+1)(2l'+1)i^{l-l'}  P_l \l \cos \theta \r P_{l'} \l \cos \theta \r F_l(k, d_X^{(2)}, r) F^{*}_{l'}(k, d_X^{(2)}, r).
\eea
Each product $P_l(x)P_{l'}(x)$ can be written as a sum of Legendre polynomials \cite{Carlitz} 
\bea
P_l(x)P_{l'}(x) &=& \sum_{r= 0}^{\text{min}(l,l')} \frac{E_r E_{l-r} E_{l'-r}}{E_{l+l' - r}} \frac{2 l + 2l' - 4r +1}{ 2 l + 2l' - 2r +1} P_{l + l' - 2r}(x) \\
&=& \sum_{m= \vert l-l' \vert}^{l+l'} \frac{1+(-1)^{\vert l-l' \vert+m}}{2}\frac{E_{\frac{l+l'-m}{2}} E_{\frac{l-l'+m}{2}} E_{\frac{l'-l+m}{2}}}{E_{\frac{l+l'+m}{2}}} \frac{2 m +1}{ l + l' + m +1} P_{m}(x)  \\
&\equiv& \sum_{m= \vert l-l' \vert}^{l+l'} C_{l,l',m} P_{m}(x) 
\eea
where $E_r = (2r-1)!!/r!$, and the third line defines $C_{l,l',m}$ for ($l$, $l'$, $m$)  that satisfy $l+ l' \geq m \geq \vert l-l' \vert$. For other combinations of these coefficients $C_{l,l',m} \equiv 0$. 
By performing the sums over  $l$ and $l^{\prime}$ in Eq.~(\ref{eq:psisqApp}), we can write 
\bea \label{eq: psi2 Legendre}
\vert \psi \vert^2 = \vert \psi_0 \vert^2 \sum_{m} c_{m}(k, d_X^{(2)}, r) P_{m} \l \cos \theta \r, 
\eea
where we have defined
\bea
\label{eq:cldef}
c_{m}(k, d_X^{(2)}, r) \equiv \sum_{l,l'} (2l+1) (2l'+1) i^{l-l'} C_{l,l',m} F_l(k, d_X^{(2)}, r) F^{*}_{l'}(k, d_X^{(2)}, r).
\eea
 For the first few values of $m$ we have 
\bea
c_{0}(k, d_X^{(2)}, r) &=& \sum_{l} (2l+1)  \vert F_l \vert^2  \\
c_{1}(k, d_X^{(2)}, r) &=& \sum_{l} 3 (l+1) \text{Im} (F_l F^{*}_{l+1} )\\
c_{2}(k, d_X^{(2)}, r) &=& \sum_{l} \left\lbrace \frac{5 l (l+1)(2l+1)}{(2l+3)(2l-1)} \vert F_l \vert^2 - \frac{15 (l+2)(l+1)}{(2l+3)} \text{Re} (F_l F^{*}_{l+2} )\right\rbrace,
\eea
where we have suppressed the arguments of $F_l(k, d_X^{(2)}, r)$ for brevity. We can use the general form from Eq. \eqref{eq: psi2 Legendre} to perform an integration over the momentum distribution
\bea
\phi^2(x) &=& \int dk \, d \cos \theta_k \, d \varphi_k k^2 f(k, \cos \theta_k)  \vert \psi \vert^2  \\
&=& \sum\limits_l \int dk \,k^2 c_l (k,d_X^{(2)}, r) \int d \cos \theta_k   f(k, \cos \theta_k)  \int d \varphi_k P_{l} (\cos \theta) 
\eea
Since only $\cos \theta$ depends on $\varphi_k$, see Eq.~(\ref{eq:dotproduct}), we perform this integration first, which results in a useful factorization:
\bea
\int\limits_{0}^{2\pi} d \phi_k P_l (\cos \theta) = 2 \pi P_l (\cos \theta_k) P_l (\cos \theta_r).
\eea
This can be used to extract the coefficients of the multipole expansion in $\cos \theta_r$ by performing integrals over the momentum amplitude $k$ and its polar coordinate $\cos \theta_k$:
\bea
\phi^2(x)
&=& 2\pi  \sum\limits_l P_l (\cos \theta_r) \int dk \, k^2 c_l (k,d_X^{(2)}, r) \int d\cos {\theta_k}\,    f(k, \cos \theta_k)  P_{l} (\cos \theta_k).  %\\ &\equiv& \sum_l a_l (k_0, d_X^{(2)}, r) P_l (\cos \theta_r).
\eea
The integral over $\cos \theta_k$ can be performed analytically. The result is
\bea
f_l(k) &\equiv& \int\limits_{-1}^{1} d \cos {\theta_k } \,  f(k, \cos \theta_k)  P_{l} (\cos \theta_k) \label{eq:fldef}\\
&=& \l \frac{1}{2 \pi \sigma_k^2} \r^{3/2} e^{ -\frac{k^2+k_0^2}{2 \sigma_k^2} } \left. 2^l \sum_{m=0}^l \binom{l}{m} \binom{\frac{l+m-1}{2}}{l} \l\frac{\partial}{\partial b}\r^m \frac{2 \sinh b}{b}\right\vert_{b = \frac{k k_0}{ \sigma_k^2}}. \nonumber 
\eea
Recalling the definition of $a_{l}$ through
\begin{equation}
    \phi^{2}(x) \equiv \sum_l a_l (k_0, d_X^{(2)}, r) P_l (\cos \theta_r),
\end{equation}
we have 
 a master expression for $a_l$, which can be used in calculations, with $f_{l}$ defined in Eq.~(\ref{eq:fldef}) and $c_{l}$ given in Eq.~(\ref{eq:cldef}):
\bea
a_l (k_0, d_X^{(2)}, r) = 2\pi   \int dk k^2 c_l (k,d_X^{(2)}, r) f_l(k).
\eea
As examples, we show expressions for $a_0$ and $a_1$: 
\bea
\label{eq:Appa0}
a_0 =  \frac{1}{\sqrt{2\pi} k_0 \sigma_k } \int dk k \l e^{ - \frac{(k-k_0)^2}{2 \sigma_k^2}} -e^{ - \frac{(k+k_0)^2}{2 \sigma_k^2}} \r \sum_{l}  (2l+1)\vert F_{l}(k, d_X^{(2)}, r) \vert^2; 
\eea

\bea
a_1 = - \frac{6}{\sqrt{2\pi} k_0 \sigma_k } \int dk  \l \l k + \frac{\sigma_k^2}{k_0} \r e^{- \frac{(k+k_0)^2}{2 \sigma_k^2}} + \l k-\frac{\sigma_k^2}{k_0} \r e^{- \frac{(k-k_0)^2}{2 \sigma_k^2}} \r \nonumber \\ \times \sum_{l}  (l+1)\text{Im}(F_{l+1}(k, d_X^{(2)}, r) F^{*}_{l}(k, d_X^{(2)}, r)).
\label{eq:Appa1}
\eea

\subsection{Example: exterior solution for scattering off a sphere}
To demonstrate the utility of this method, we apply it to the exterior solution where $F_l = j_l (kr) + A_l h_{l}(kr)$. The first term results from projecting a plane wave into spherical waves, while the second term arises from the scattered wave.
While $A_l$ vanishes for $l \gg k R_\oplus$, the term proportional to $j_{l}(kr)$ is suppressed when $l \gg k r$. This indicates that the evaluation of Eqs.~(\ref{eq:Appa0}),(\ref{eq:Appa1}) would require the inclusion of terms in the $l$ sum up to $\sim k r$, which for $r\gg R_{\oplus}$ could present a numerical challenge.  Fortunately, since the plane wave on its own gives rise to a uniform field profile ($a_0 = 1$ and $a_{i>0} = 0$), we can trivially sum the terms proportional to $j_l^2(kr)$.  The remaining terms (proportional to $A_{l})$ converge at $l\sim k R_\oplus$. For the first two $a_l$ coefficients we get
\bea
a_0 = 1 +  \frac{1}{\sqrt{2\pi} k_0 \sigma_k } \int dk k \l e^{ - \frac{(k-k_0)^2}{2 \sigma_k^2}} -e^{ - \frac{(k+k_0)^2}{2 \sigma_k^2}} \r \\
\times
\sum_{l}  (2l+1) \l 2 \text{Re}(A_l j_l (kr) h_l(kr)) + \vert A_l h_l (kr) \vert^2  \r,  
\nonumber 
\eea

and

\bea
a_1 = &-& \frac{6}{\sqrt{2\pi} k_0 \sigma_k } \int dk  \l \l k + \frac{\sigma_k^2}{k_0} \r e^{- \frac{(k+k_0)^2}{2 \sigma_k^2}} + \l k-\frac{\sigma_k^2}{k_0} \r e^{- \frac{(k-k_0)^2}{2 \sigma_k^2}} \r  \\ &\times& \sum_{l}  (l+1)\text{Im}( A^{*}_{l+1} j_l(kr) h^{*}_{l+1}(kr) + A_{l} j_{l+1}(kr) h_{l}(kr) + A_{l} A^{*}_{l+1} h_{l}(kr) h^{*}_{l+1}(kr) ).  \nonumber 
\eea

\bibliographystyle{JHEP}
\bibliography{biblio}{}

\end{document}